\documentclass[preprints,article,accept,moreauthors]{mdpi} 
\firstpage{1} 
\pubvolume{1}
\issuenum{1}
\articlenumber{0}
\pubyear{2026}
\copyrightyear{2026}
\datereceived{ } 
\dateaccepted{ } 
\datepublished{ } 
\hreflink{https://doi.org/} 

\usepackage{graphicx}	
\usepackage{multirow}
\usepackage{todonotes}
\usepackage{multicol}
\usepackage{marginnote}
\usepackage{aas_macros}

\newcommand{\jetcaf}{{\fontfamily{qcr}\selectfont JeTCAF}}
\newcommand{\tcaf}{{\fontfamily{qcr}\selectfont TCAF}}

\newcommand{\tbabs}{{\fontfamily{qcr}\selectfont TBABS}}

\newcommand{\pcfabs}{{\fontfamily{qcr}\selectfont PCFABS}}

\newcommand{\gauss}{{\fontfamily{qcr}\selectfont GAUSS}}

\Title{15-Year X-ray Study Links Variable Mass Accretion to Changing Look Behavior in AGN NGC 7582}

\Author{Santanu Mondal $^{1,}$*\orcidA{}, V. Jithesh $^{1}$, Neeraj Kumari $^{2}$ and Tek P. Adhikari $^{3,4}$}

\AuthorNames{Santanu Mondal, V. Jithesh, Neeraj Kumari and Tek P. Adhikari}

\address{%
$^{1}$ \quad Department of Physics and Electronics, Christ University, Bangalore 560029, Karnataka, India; santanuicsp@gmail.com\\
$^{2}$ \quad INAF-IASF, Via Ugo La Malfa, 153, Palermo, I-90146, Italy\\
$^{3}$ \quad CAS Key Laboratory for Research in Galaxies and Cosmology, Department of Astronomy, University of Science and Technology of China, Hefei, Anhui 230026, China\\
$^{4}$ \quad School of Astronomy and Space Science, University of Science and Technology of China, Hefei, Anhui 230026, China}

\corres{Correspondence: santanuicsp@gmail.com}

\abstract{Changing Look Active Galactic Nuclei (CLAGN) are a subclass of AGN that exhibit spectral state transitions in both optical/UV and X-ray bands on timescales of days to years. In this work, we studied fifteen years of X-ray data of the CLAGN NGC 7582 taken from {\it Suzaku}, {\it XMM-Newton}, and {\it NuSTAR} to understand the origin of the CL behavior. Our study reveals that the obscuration (Hydrogen column density, $N_H$) along the line of sight is associated with the failed wind {\color{black}or obscurer} scenario. The variation in mass accretion is the key that can solely explain the origin of both the failed wind and CL behavior. The mass accretion behavior in NGC 7582 resembles the hard spectral state of low mass X-ray binaries. Furthermore, the observed CL timescales in NGC 7582 can be explained either by the dynamical timescale at the BLR or by viscous and thermal timescales at the boundary layer of the corona for the inferred black hole mass of $5.3^{+0.5}_{-0.3} \times 10^7 M_\odot$. Our study thus highlights a direct physical link between the corona and the BLR, which may be potential drivers of the CL behavior.} 


\begin{document}

\setcounter{section}{0}
\section{Introduction}
Active Galactic Nuclei (AGN) host supermassive black holes (SMBHs) at the center, which are believed to be accreting actively and leaving imprints in both their spectral and timing properties. Such sources change their spectral states on a timescale of hours to decades, depending on the underlying physical processes driving them. One such class of AGN is `changing look' (CL) AGN, which experiences rapid changes in its optical, ultraviolet, or X-ray classifications. In the X-ray band, these changes can be mostly attributed to changes in the line-of-sight (LOS) column density ($N_H$), known as Changing-obscuration \citep[CO-AGN,][and references therein]{Risalitietal2005,Riccietal2016}, while in the optical/UV band these events are usually associated with the changes in the accretion-driven radiation field, leading to the changes in optical/UV line profile, known as Changing-state \citep[CS-AGN][for a review]{Riccietal2020,Ricci2023NatAs...7.1282R}. In the X-ray band, the evolution of relativistic outflows and the corona, along with a 1 keV emission line in the spectrum, can also indicate CL behavior, as observed in 1ES 1927+654 \citep{MastersonEtal2022ApJ...934...35M}. Additionally, some AGN exhibit significant variations in the X-ray continuum luminosity that occur in tandem with the optical/UV CL behavior (Sy~1 to Sy~2 and vice versa)\citep{LaMassa2017,Kollatschny2020,Guolo2021}.
In this context, \citet{AdhikariEtal2025ApJ...988...25A} showed that
coordinated changes in the accretion-disk and X-ray continuum can account for the observed variability of optical/UV broad emission lines in CLAGN. A series of studies attempted to understand the physical drivers of such change in states \citep[][and references therein]{NodaDone2018,Ricci2023NatAs...7.1282R,CaglarEtAl2023ApJ...956...60C,TempleEtal2023MNRAS.518.2938T,YamadaEtAl2024ApJS..274....8Y}. However, the physical origin is not clear to date; associating the underlying reason is driven by variation in mass accretion rate. Recent studies revealed that the physical origin of both states is interlinked through the accretion-wind scenario \citep{MondalEtal2022A&A...662A..77M,LyuEtal2025MNRAS.537.1099L}. A growing number of studies have reported that the mass accretion rate variation can be the primary driver for the CL phenomena \citep[][and references therein]{MacLeodEtal2019ApJ...874....8M,MondalEtal2022A&A...662A..77M,SwainEtal2023MNRAS.520.3712S,TempleEtal2023MNRAS.518.2938T,JanaEtAl2025A&A...693A..35J,TitarchukEtAl2025A&A...693A.126T}.

NGC 7582 (J2318-4222) is a nearby (z = 0.00525) X-ray bright Seyfert 2 galaxy with SMBH mass $3.6-8.1 \times 10^7$ M$_\odot$ \citep{WoldEtal2006A&A...460..449W}. The galaxy is located at Galactic latitute $(l)$ $348.08^{\circ}$ and longitude $(b)$ $-65.69^{\circ}$ and disk inclination $58^{\circ} - 65^{\circ}$ \citep{WoldEtal2006A&A...460..449W}. The nucleus of this galaxy showed
significant spectral variation in X-ray, which helps understand the dynamical environment and the possible accretion behavior, including the geometry, inflow-outflow properties, and the supermassive black hole (SMBH) intrinsic parameters such as mass and spin. 
Early X-ray observations of NGC 7582 by HEAO-1/A2 \citep{Mushotzky1982ApJ...256...92M}, Einstein Observatory \citep{Maccacaro1981ApJ...246L..11M,ReichertEtal1985ApJ...296...69R}, EXOSAT \citep{TurnerPound1989MNRAS.240..833T}, Ginga \citep{WarwickEtal1993MNRAS.265..412W}, ASCA \citep{SchachterEtal1998ApJ...503L.123S,XueEtal1998PASJ...50..519X}, and BeppoSAX \citep{TurnerEtal2000ApJ...531..245T} revealed a very flat observed spectrum with a signature of a full absorption layer with $N_H$ from $\sim 10^{22}$ to $\sim 10^{24}$ cm$^{-2}$ \citep[][and references therein]{BianchiEtal2009ApJ...695..781B,BraitoEtal2017A&A...600A.135B,LefkirEtal2023MNRAS.522.1169L} with at least a few CO transitions. In addition to the CO event, NGC 7582 showed a rapid transition from a Type 2 Seyfert to Type 1 in a timescale of 20 days, with a broad H$\alpha$ followed by a slow decay of the broad lines, with a possibility of a CL event, where the obscuring material blocking the broad lines moved temporarily out of the line of sight \citep{AretxagaEtal1999ApJ...519L.123A}. Several follow-up X-ray studies since then inferred this source as a CLAGN. 

The results from the {\it XMM-Newton} data conducted during 2001 and 2005 \citep{PiconcelliEtal2007A&A...466..855P} suggested that the spectrum of the continuum emission is very complex. It can be well described by a model consisting of a combination of a heavily absorbed power law and a pure reflection component, with a significant increase in the column density between 2001 and 2005. During this period, the lowest flux was detected in comparison to the 1998 {\it BeppoSAX} observation, indicative of a drastic spectral state change \citep{PiconcelliEtal2007A&A...466..855P}. Later, spectral modeling to describe four {\it Suzaku} observations from 2007 with an additional {\it XMM-Newton} observation from 2007 found spectral variability on timescales from $<1$ day to 7 months and concluded that it can be due to variable absorption of the continuum power law by hidden broad line region (BLR) clouds \citep{BianchiEtal2009ApJ...695..781B}. \citet{RiversEtal2015ApJ...815...55R} analyzed the {\it NuSTAR} data of the source during 2012 and reported the presence of a patchy torus that heavily absorbed the continuum and an association of multiple absorbing layers. After the inclusion of more epochs of data, \citet{BalokovicEtal2018ApJ...854...42B} obtained a similar covering factor and absorbing column density values; however, it challenged the general unified model of AGN based on diverse structures of the torus. Very recently, Lefkir et al. \citep{LefkirEtal2023MNRAS.522.1169L}, analyzed both {\it XMM-Newton} and {\it NuSTAR} data and put constraints on the variable cloud properties along the LOS of the source. The cloud column density varies from Compton-thin to Compton-thick between observations. Moreover, a dust lane along the line of sight can provide additional obscuring material, leading to substantial coverage of the central AGN \citep{BianchiEtal2007MNRAS.374..697B}. NGC 7582 shows clear ionized gas outflows along its kpc-scale ionization cone, with blue-shifted velocities up to a few hundred km s$^{-1}$ \citep{MorrisEtal1985MNRAS.216..193M,RiffelEtal2009MNRAS.393..783R}. High-resolution radio observations reveal extended emission that may trace a weak AGN jet or AGN-driven strong outflows \citep{ForbesNorris1998MNRAS.300..757F,OrientiPrieto2010MNRAS.401.2599O}. Recent {\it JWST} study further strengthens that possibility \citep{VeenemaEtal2025MNRAS.544.3361V}. Therefore, there can be multiple absorbers at different length scales.

Therefore, extensive spectral studies of NGC\,7582 in X-rays and also in other bands are done to understand the gas properties in the circumnuclear region of the central SMBH. Such studies probe the radiation mechanisms involved in the vicinity of the SMBH and the origin of the observed variabilities. However, none of these studies considered the accretion-ejection properties and flow configuration to directly model the data to explain the CL phenomena in NGC 7582. Strong outflows in NGC 7582 in different length scales have been detected by multi-instruments as discussed above, which motivated us to use accretion-ejection-based \jetcaf\, \citep{MondalChakrabarti2021} or Jet in Two Component Advective Flow (\tcaf)  model for this source, in place of \tcaf\,model \citep{ChakrabartiTitarchuk1995}. The \jetcaf\,model is an upgraded version of the \tcaf\,model, where we have incorporated jet or outflow component, which comes from the shock region. In this model, there are three components: 1. A cold Keplerian component at the equatorial plane of the accretion disk, 2. Low angular momentum sub-Keplerian components above and below the Keplerian component, which moves supersonically as the matter comes closer to the BH and forms a shock following Rankine-Hugoniot jump conditions \citep{Chakrabarti1989}, and 3. The jet or outflow component, which is another Comptonizing medium for both the corona and disk photons. As the flow satisfies the R-H condition, a shock forms, which generates heat that puffs up the inner disk as corona and drives some fraction of the inflowing matter as outflows \citep{Chakrabarti1999}. Therefore, the disk truncates at the shock location, which also behaves as the boundary layer of the corona. We used all three components and their corresponding contribution in the emitted spectrum as discussed later in Appendix \ref{sec:apenA}. In this work, we have analyzed 15 years of X-ray data of the source NGC 7582 to understand the CO behavior and the origin of drastic variability in spectral states of the source from a physically motivated \jetcaf\,model. The paper is organized as follows: the observations and data analysis procedures are discussed in the next section, in Sec 3, we discuss the spectral fitting and results, and finally, we draw our conclusion in Sec 4.

\section {Observations and data analysis}\label{sec:Observation}

We used 15 years of data from {\it XMM-Newton}, {\it Suzaku}, and {\it NuSTAR} of NGC\,7582 from 2001 to 2016. The observation log is given in Table \ref{table:observation}.

{\it \textbf{XMM-Newton:}} For this study, we used the European Photon Imaging Camera (EPIC)-pn \citep{Jansen2001} due to its higher sensitivity in the energy range 0.3-10 keV. All observations used in this work have been taken in full-frame imaging mode. The EPIC-pn data is reprocessed using standard \textsc{Science Analysis System (SAS v.18.0.0;}) software and updated calibration files. To check the presence of any flaring particle background during the observations, we generated a light curve above 10 keV and created good time interval (GTI) files by excluding the time intervals containing particle background, which were capped by count rate. We considered events with {\tt PATTERN}$\le$4 for EPIC-pn. The observed data are checked for the photon pile-up effect using the {\tt epatplot} task. We extracted source and background spectra by selecting circular regions of 30 arcsec radius. The background region is selected from the same CCD where no other X-ray source was present. The redistribution matrix and ancillary response files were generated by {\tt arfgen} and {\tt rmfgen} tasks, respectively. We have not used the observation conducted in April 2007 due to its low exposure ($\sim 2$ ks) and noisy spectrum.

\begin{table}
\small
\centering
	\caption{\label{table:observation} Log of observations of NGC\,7582. The $^\ast$ denotes simultaneous observations. The corresponing energy band for each instrument is also given.}
\begin{tabular}{cccccccc}
	Obs. Id. &Exp. & Epoch  & Date   & MJD & Energy range\\
	   &(ks) &        &        &    & (keV)\\
\hline
&&{\it XMM-Newton}&&\\
	0112310201&23 &X1&2001-05-25&52054&0.3-10 \\
	0204610101&102&X2&2005-04-29&53489& \\
	0782720301&101&X3&2016-04-28&57506$^\ast$& \\
&&{\it Suzaku}&&\\
	702052010&23.9&S1&2007-05-01&54221&0.6-10  \\
	702052020&28.8&S2&2007-05-28&54248&  \\
	702052030&29.4&S3&2007-11-09&54413&  \\
	702052040&31.9&S4&2007-11-16&54420& \\
           &&{\it NuSTAR}&&\\
	60061318002&16.5&N1&2012-08-31&56170& 3-75 \\
	60061318004&14.6&N2&2012-09-14&56184& \\
	60201003002&48.5&XN3&2016-04-28&57506$^\ast$& \\

\hline
\end{tabular}
\end{table}

{\it \textbf{Suzaku:}} The source was observed by {\it Suzaku} on four occasions, and details are given in Table \ref{table:observation}. The unﬁltered {\it Suzaku} data were downloaded from the HEASARC website and reprocessed the X-ray imaging spectrometer (XIS) data using the speciﬁc HEADAS tool {\sc AEPIPELINE}. The source events were extracted from a circular region of radius 220 arcsec in the three detectors (XIS0, XIS1 and XIS3). We selected two circular regions of radius 110 arcsec near the source region and extracted the background events. The front illuminated (FI) CCD spectra, XIS0
and XIS3, were added using the FTOOL ADDASCASPEC.
The co-added spectra were then grouped, with a minimum of 20 counts per bin, for spectral modeling. We fitted the co-added FI spectrum in the 0.6 - 10 keV energy band, and removed the 1.7-2.0 keV energy range due to calibration issues. 

{\it \textbf{NuSTAR:}} The {\it NuSTAR} data were extracted using the standard 
{\sc NUSTARDAS v1.3.1}
software. The {\tt nupipeline} task is used to generate cleaned event lists and 
{\tt nuproducts} to generate the spectra. A source region of 
$50^{\prime\prime}$  and a background region of $80^{\prime\prime}$ are used in {\sc ds9} to extract the source and background spectra. The data were grouped with a 
minimum of 20 counts in each bin using {\tt grppha} task. The spectral modeling was done for the individual epochs and for the joint epochs where applicable. 


We used {\sc XSPEC} 
\citep{Arnaud1996} version 12.12.1 for spectral analysis. 
The broadband energy range 0.3-75 keV covered by the spectra of NGC\,7582 showed complex features including Fe K$\alpha$ line, Compton hump above 10 keV, soft-excess, and absorption. Each epoch of observation was fitted using an accretion-ejection-based \jetcaf\,model that has been expanded to include jet or outflow emission in \tcaf\,model. The jet or outflow is extended till the sonic point, which is $\sim 2-2.5$ times the size of the corona \citep{Chakrabarti1999}. We note that the sonic point signifies the boundary where the primary outflow driving mechanisms can successfully overcome gravitational binding and inertia, allowing the material to transition into a freely streaming, high-velocity outflow. The Compton hump above 10 keV was taken care of through the Comptonization of hard photons by cold disk material similar to \tcaf\,model. An additional hump in the spectrum originates from the scattering of the Comptonized photons by the jet/outflows in the \jetcaf\, model. We note that \jetcaf\,is not a specific model for jetted sources, but rather a general model; it has the flexibility in collimation parameter that can fit both collimated and isotropic outflows from the corona. An illustration of the model is shown in \autoref{fig:JeTCAFIllus} and the origin of its different spectral components is discussed in Appendix \ref{sec:apenA}. The \jetcaf\,model does not have the so-called reflection from cold or ionized medium that can produce the Fe K line as well as a hump. Therefore, to fit the Fe K line at $\sim 6.4$\,keV in the spectrum, \gauss\,model is used. It provides the line energy ($E_g$), its width, and normalization or strength. However, our primary goal is not to study the line properties, but the role of accretion in CL phenomena. We used multiple \gauss\,components below 2 keV to fit the soft excess. Note, these components are not required for {\it NuSTAR} data. For the Galactic absorption, \tbabs \citep{Wilmsetal2000} is used, where the hydrogen column density is fixed to $1.5\times 10^{20}$cm$^{-2}$ \citep{Kalberlaetal2005} for all data sets. We used the partial covering absorption (\pcfabs) to model the neutral or weakly ionized obscurer, which takes into account for the clumpy, inhomogeneous distribution of obscuring gas that partially covers the LOS to the central X-ray source. The physical properties of this obscurer are characterized by two primary parameters: the equivalent hydrogen column density ($N_{\rm Hpcf}$), which traces the optical depth and density of the obscuring clouds, and the dimensionless covering factor ($C_f$), where $0 < C_f < 1$. The second parameter represents the attenuation of a fraction of the primary X-ray emission by these dense, partially ionized, or neutral clumps, while the remaining fraction ($1 - C_f$) leaks through the medium unabsorbed. The full model reads in {\it XSPEC} for both individual and joint fitting as \tbabs*\pcfabs(\gauss+\jetcaf).

\begin{table*}
\scriptsize
\centering
\caption{\label{table:JetcafResults} Best-fitted model parameters using \tbabs*\pcfabs(\gauss+\jetcaf) Here, $\dot m_d$, $\dot m_h$, $X_s$, $R$, and $f_{\rm col}$ are the disk and halo mass accretion rates, location of the shock or size of the corona, shock compression ratio, and the jet collimation factor, respectively in \jetcaf\,model. N$_{\rm Hpcf}$ is the hydrogen column density in \pcfabs\,model. The Galactic absorption in \tbabs\,model is fixed to $1.5\times10^{20}$ cm$^{-2}$. All data fit required a \gauss\, model component for the Fe K$\alpha$ line at $\sim 6.4$ keV. The last column, $L_{3-10 keV}$, shows the X-ray luminosity in the 3-10 keV band in units of 10$^{41}$ erg s$^{-1}$ for all epochs. 
} 
\begin{tabular}{ccccccccccccc}
\hline
Model&\multicolumn{6}{c}{\jetcaf}&\multicolumn{2}{c}{\pcfabs}\\
\multirow{2}{*}{Epoch} &$M_{\rm BH}$ &$\dot m_{\rm d}$ & $\dot m_{\rm h}$ & $X_{\rm s}$ & R &$f_{\rm col}$&$N_{\rm Hpcf}$&$C_f$&$\chi^2/dof$&L$_{3-10 keV}$ \\
	     & $(10^7 M_\odot)$&$(10^{-3}\dot M_{\rm Edd})$&$(\dot M_{\rm Edd})$&$(r_{\rm S})$& & &$(10^{22}$ cm$^{-2})$& & &$10^{41}$ erg s$^{-1}$\\
\hline
\multicolumn{10}{c}{\it {\bf XMM-Newton}}\\
     X1   &$5.4\pm0.4$&$3.8\pm0.3$&$0.61\pm0.04$&$17.6\pm1.9$&$3.86\pm0.28$&$0.28\pm0.02$ &$49.6\pm3.8$&$0.91\pm0.04$&189/149&$2.54\pm0.03$ \\
     X2   &$5.4\pm0.3$&$3.1\pm0.2$&$0.62\pm0.02$&$26.7\pm2.7$&$2.33\pm0.08$&$0.40\pm0.03$ &$44.4\pm4.2$&$0.89\pm0.02$&544/451&$1.41\pm0.03$ \\
\multicolumn{10}{c}{\it {\bf Suzaku}}\\
     S1   &$5.1\pm0.4$&$5.0\pm0.3$&$0.51\pm0.03$&$23.7\pm2.1$&$4.31\pm0.31$&$0.21\pm0.04$&$32.9\pm4.1$&$0.94\pm0.02$&244/203&$3.23\pm0.07$ \\
     S2   &$5.3\pm0.4$&$4.1\pm0.3$&$0.55\pm0.03$&$22.0\pm1.8$&$2.73\pm0.19$&$0.40\pm0.02$ &$38.3\pm4.0$&$0.95\pm 0.02$&245/215&$2.67\pm0.06$ \\
     S3   &$5.1\pm0.2$&$2.7\pm0.2$&$0.74\pm0.05$&$32.1\pm2.8$&$5.51\pm0.24$&$0.29\pm0.04$&$47.8\pm3.9$&$0.88\pm0.05$&206/161&$1.79\pm0.03$ \\
     S4   &$5.3\pm0.3$&$2.9\pm0.5$&$0.53\pm0.02$&$18.1\pm2.0$&$3.57\pm0.22$&$0.41\pm0.05$ &$48.9\pm6.5$&$0.90\pm0.05$&217/172&$1.60\pm0.05$ \\
\multicolumn{10}{c}{\it {\bf NuSTAR}}\\
N1 &$5.2\pm0.5$&$3.0\pm0.2$&$0.57\pm0.05$&$30.9\pm2.9$&$4.81\pm0.41$&$0.72\pm0.05$&$36.5\pm2.7$&$0.87\pm0.11$&117/124&$2.88\pm0.04$\\

N2&$5.5\pm0.4$&$2.1\pm0.4$&$0.82\pm0.06$&$38.8\pm4.1$&$2.22\pm0.19$&$0.59\pm0.07$&$96.1\pm6.5$&$0.95\pm0.21$&92/82&$1.91\pm0.04$ \\
\multicolumn{10}{c}{\it {\bf XMM-Newton+NuSTAR}}\\
XN3 &$5.1\pm0.3$&$9.1\pm0.7$&$0.58\pm0.04$&$19.8\pm2.2$&$3.03\pm0.34$&$0.38\pm0.03$&$26.4\pm1.8$&$0.95\pm0.02$&1418/1285&$6.07\pm0.07$\\
\hline
\end{tabular} 
\end{table*}

\section{Results and Discussion} \label{sec:Results}

\subsection{Spectral analysis and accretion behavior}
\begin{figure*}
\centering{
\hspace{-1.2cm}
\includegraphics[height=6.0truecm,width=4.0truecm,angle=270]{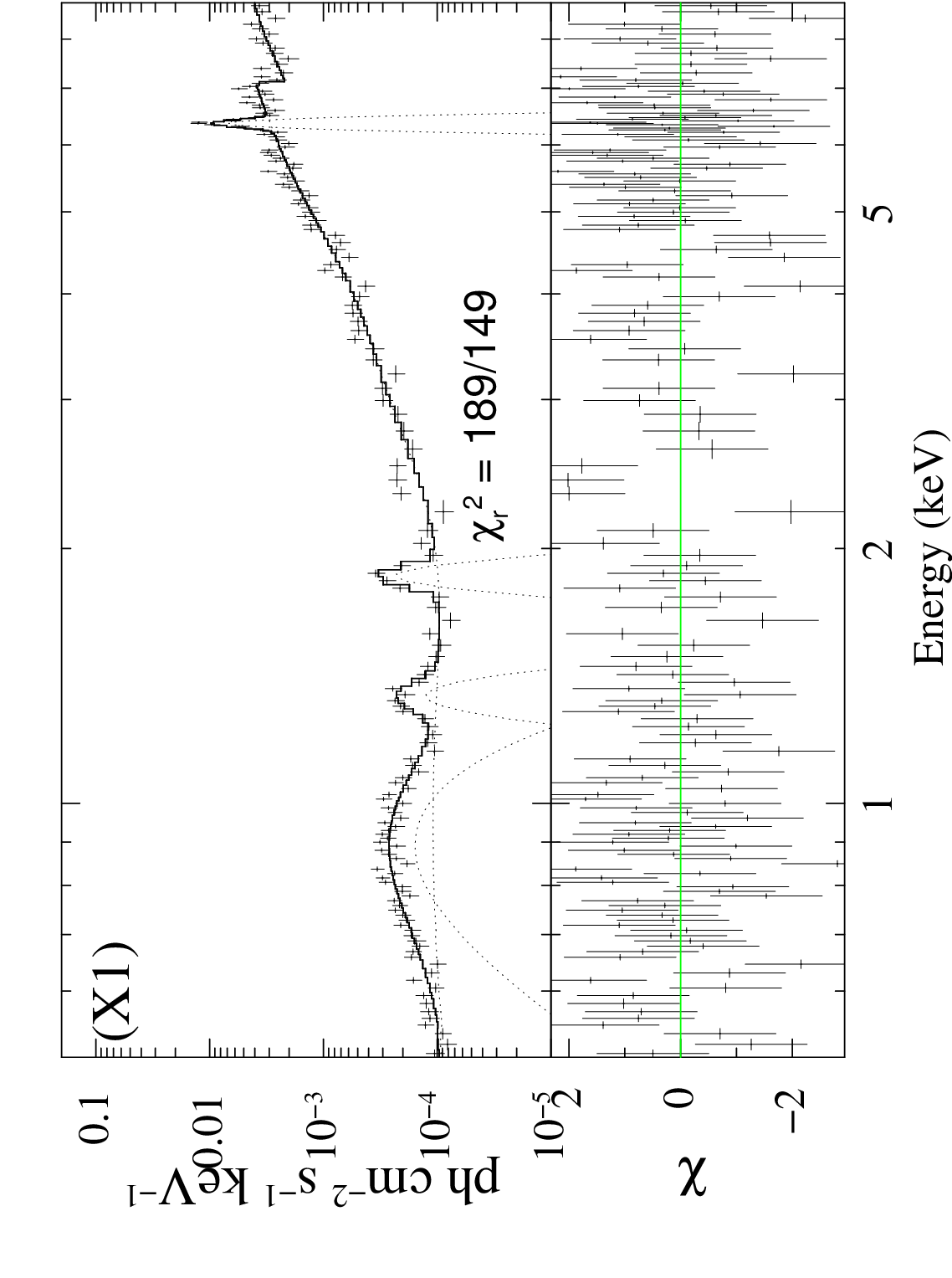}
\includegraphics[height=6.0truecm,width=4.0truecm,angle=270]{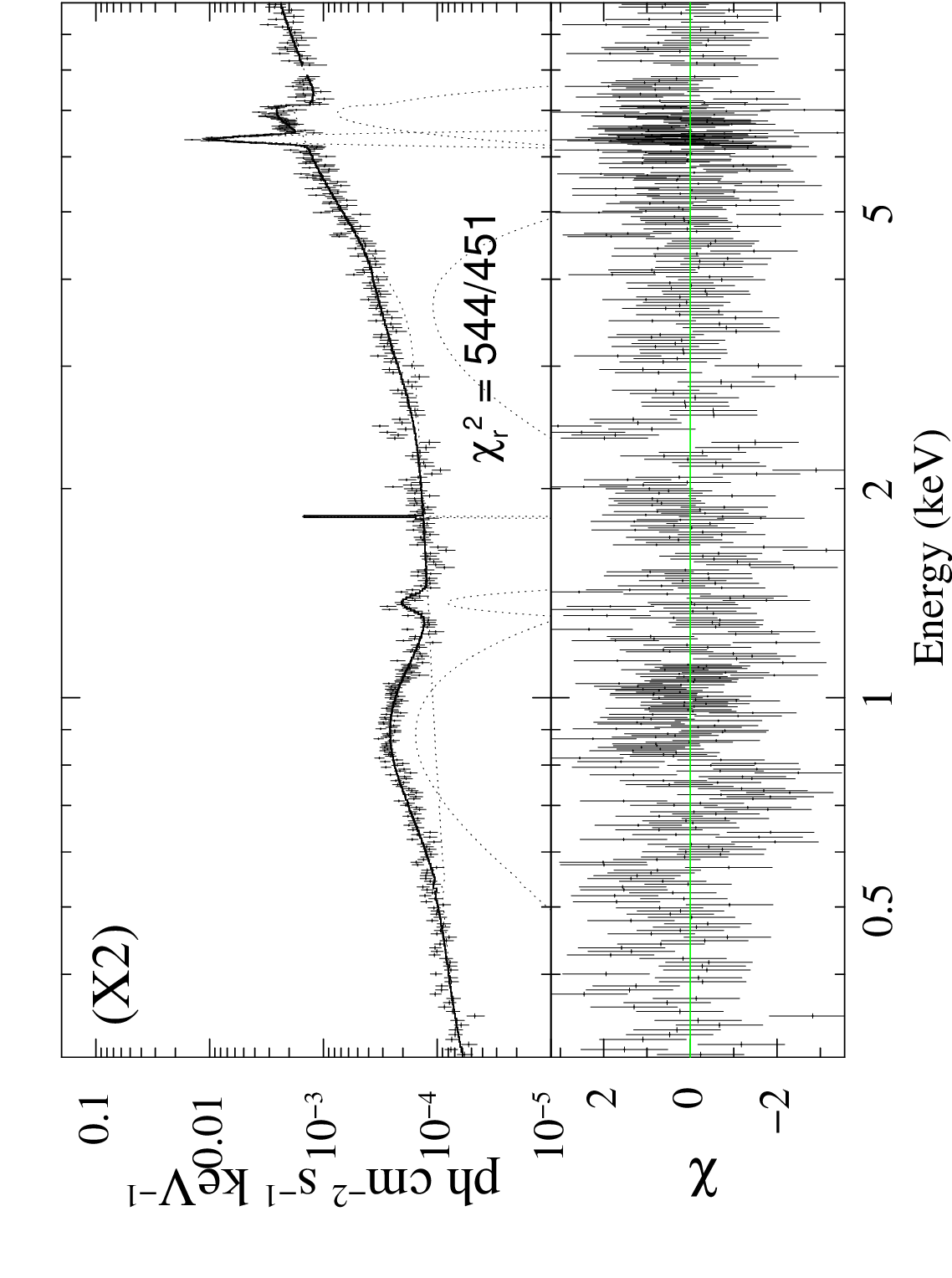}
\includegraphics[height=6.0truecm,width=4.0truecm,angle=270]{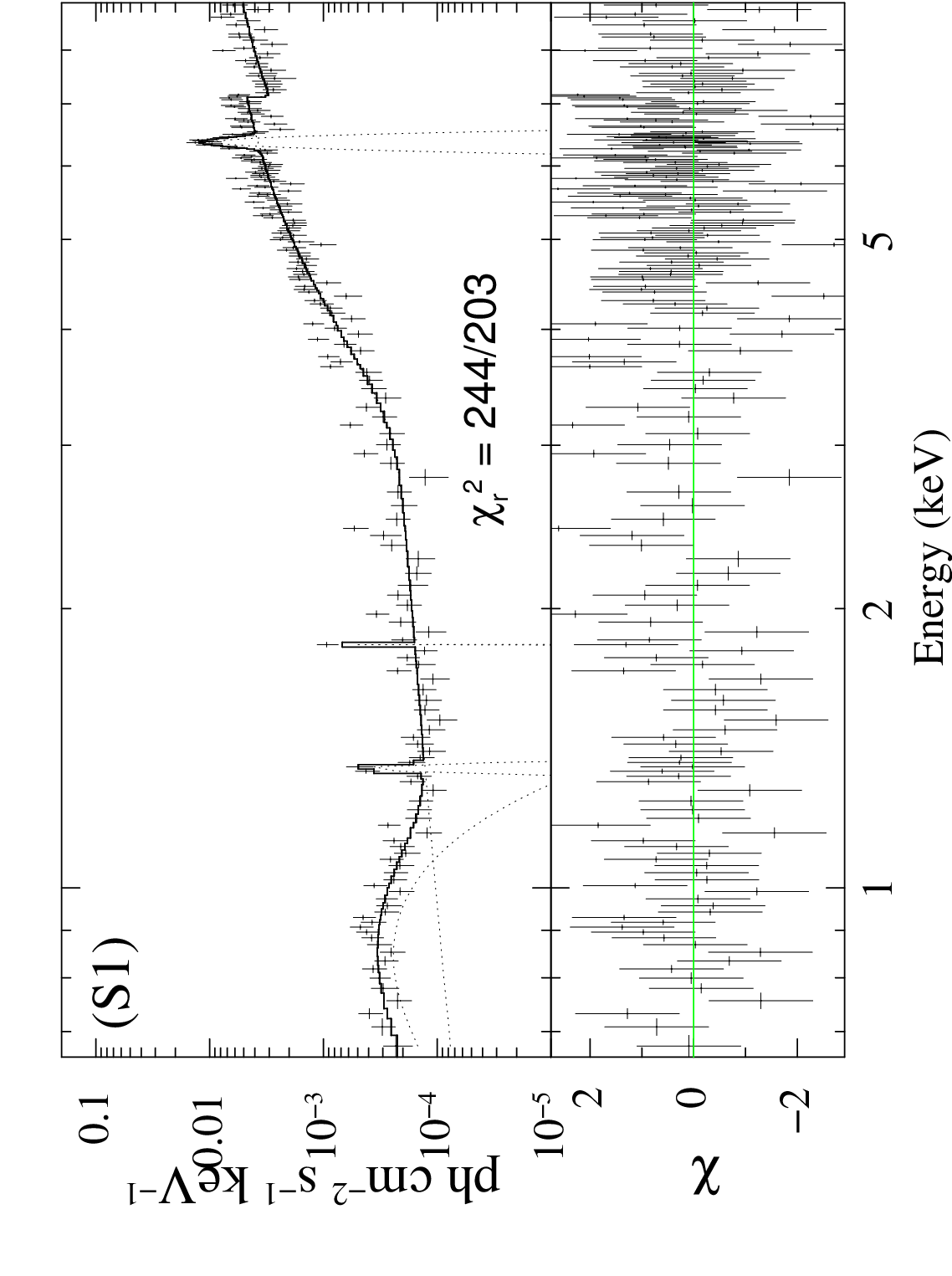}\\
\hspace{-1.2cm}
\includegraphics[height=6.0truecm,width=4.0truecm,angle=270]{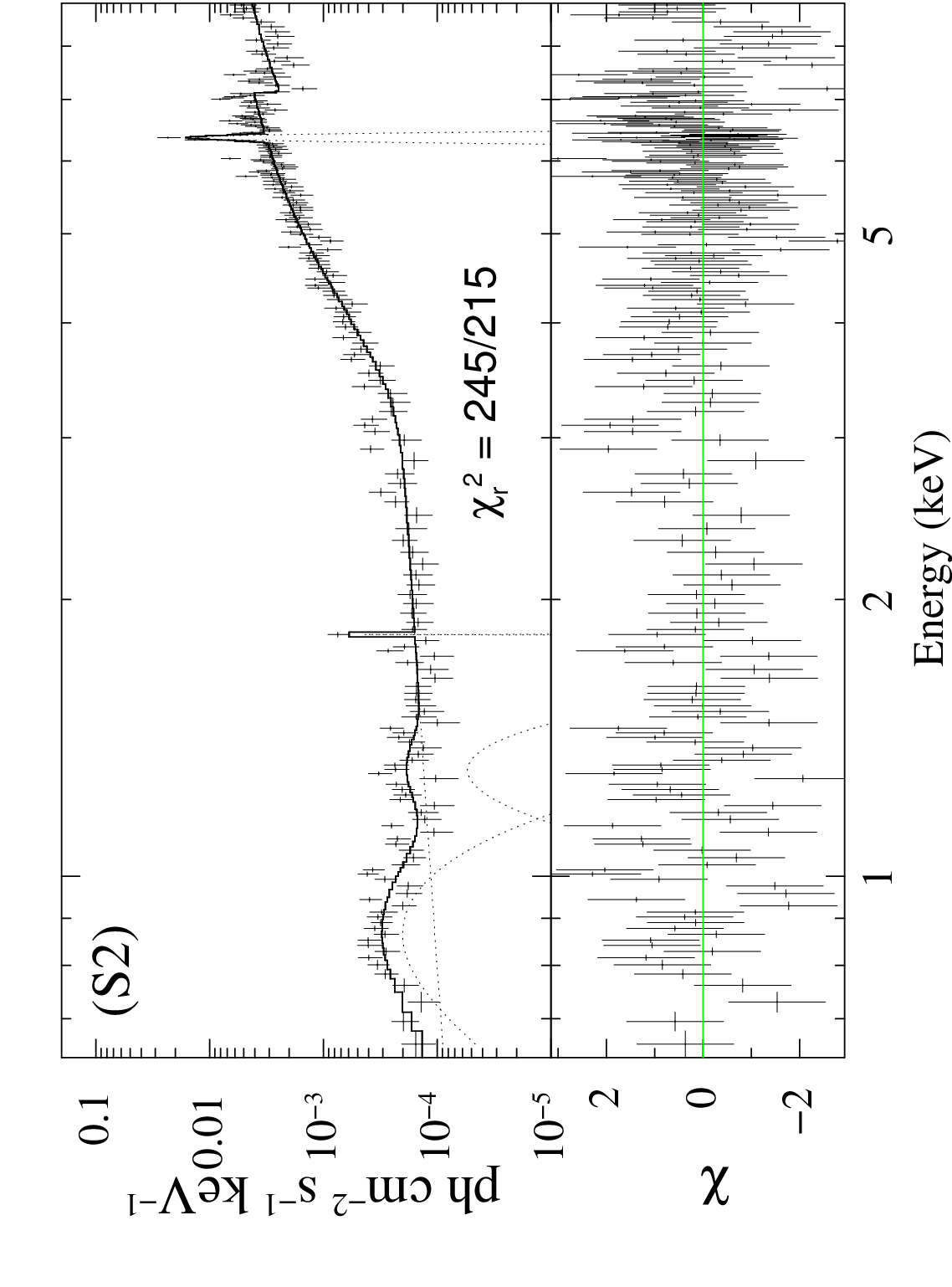}
\includegraphics[height=6.0truecm,width=4.0truecm,angle=270]{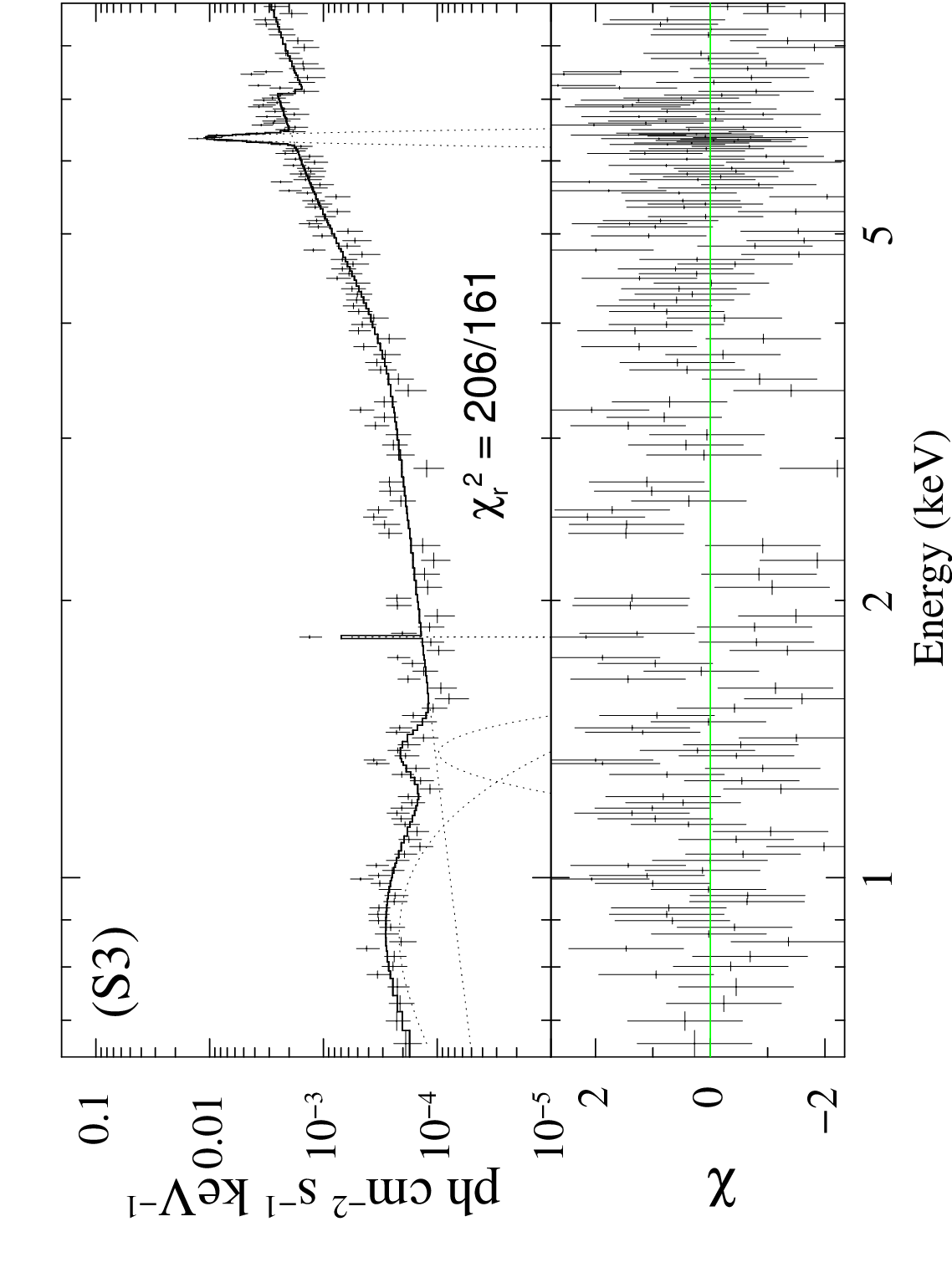}
\includegraphics[height=6.0truecm,width=4.0truecm,angle=270]{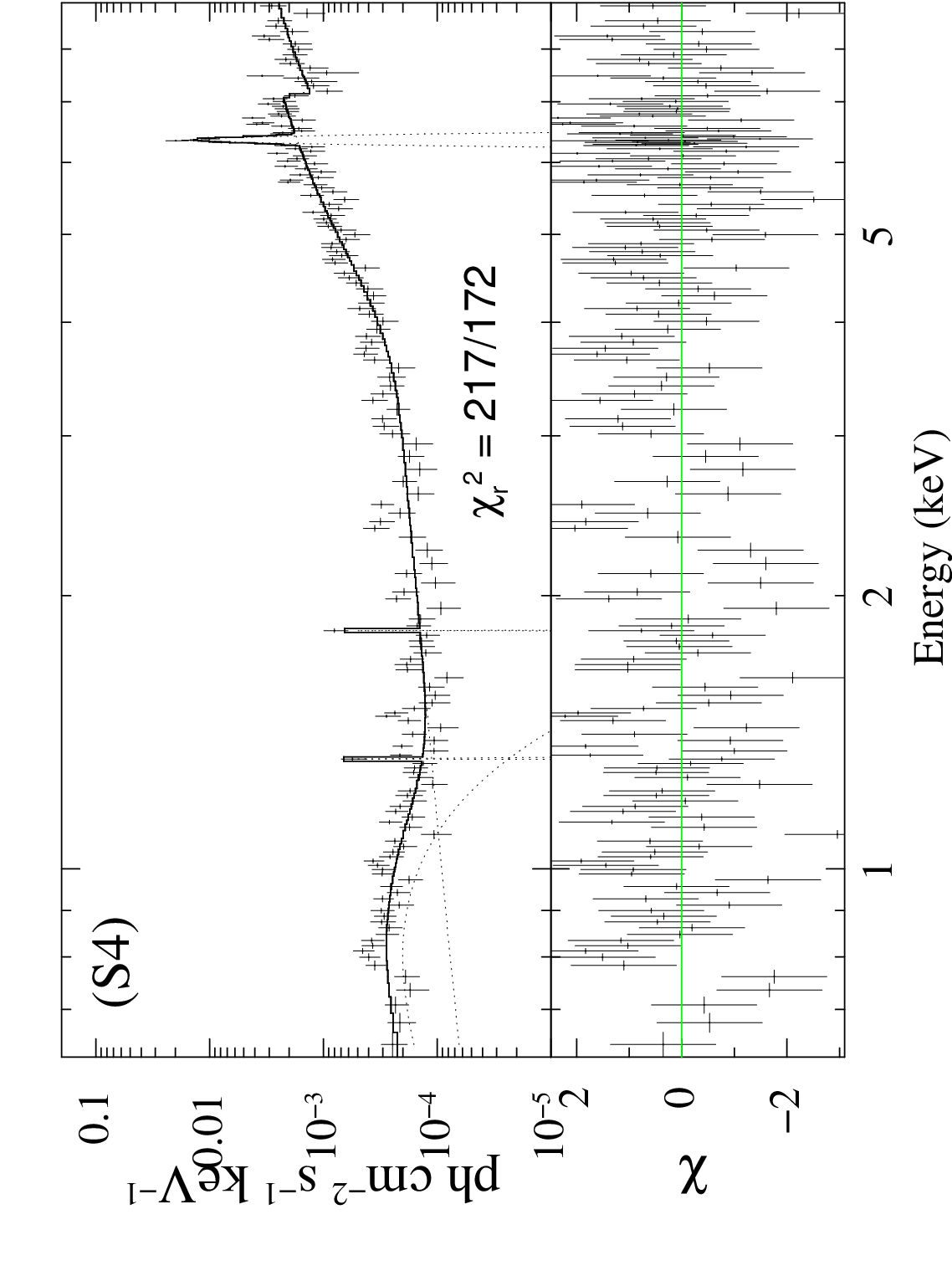}\\
\hspace{-1.2cm}
\includegraphics[height=6.0truecm,width=4.0truecm,angle=270]{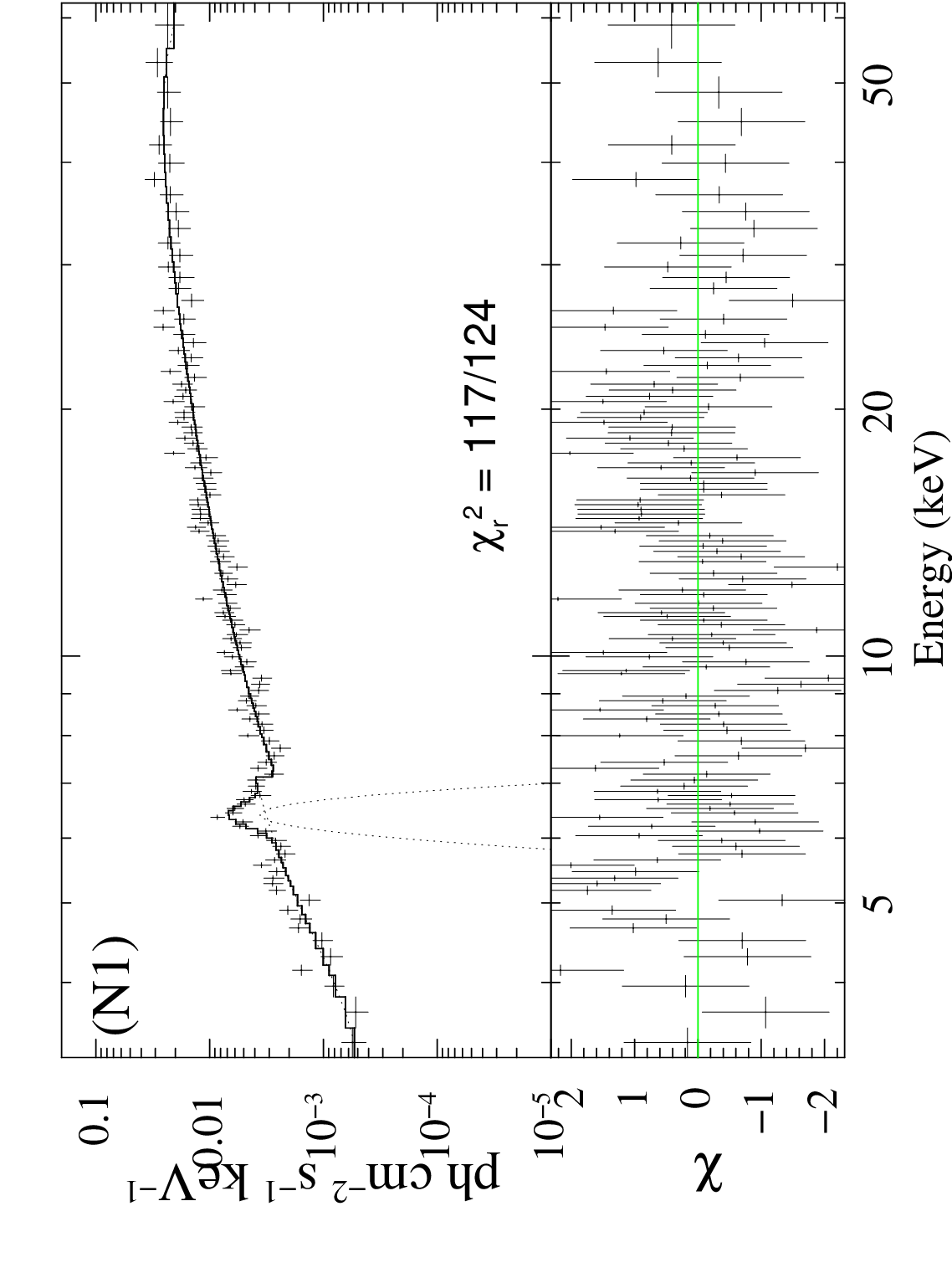}
\includegraphics[height=6.0truecm,width=4.0truecm,angle=270]{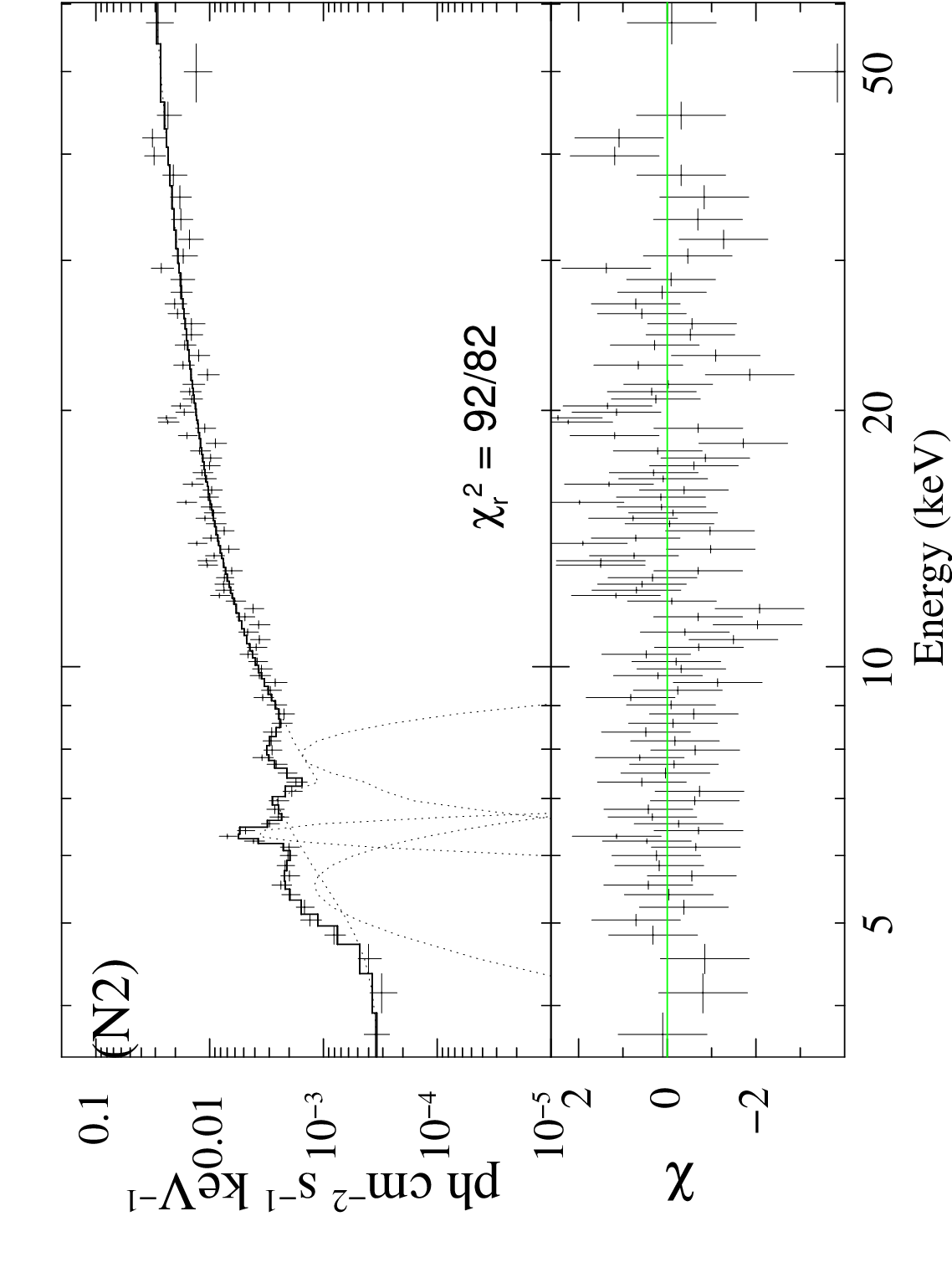}
\includegraphics[height=6.0truecm,width=4.0truecm,angle=270]{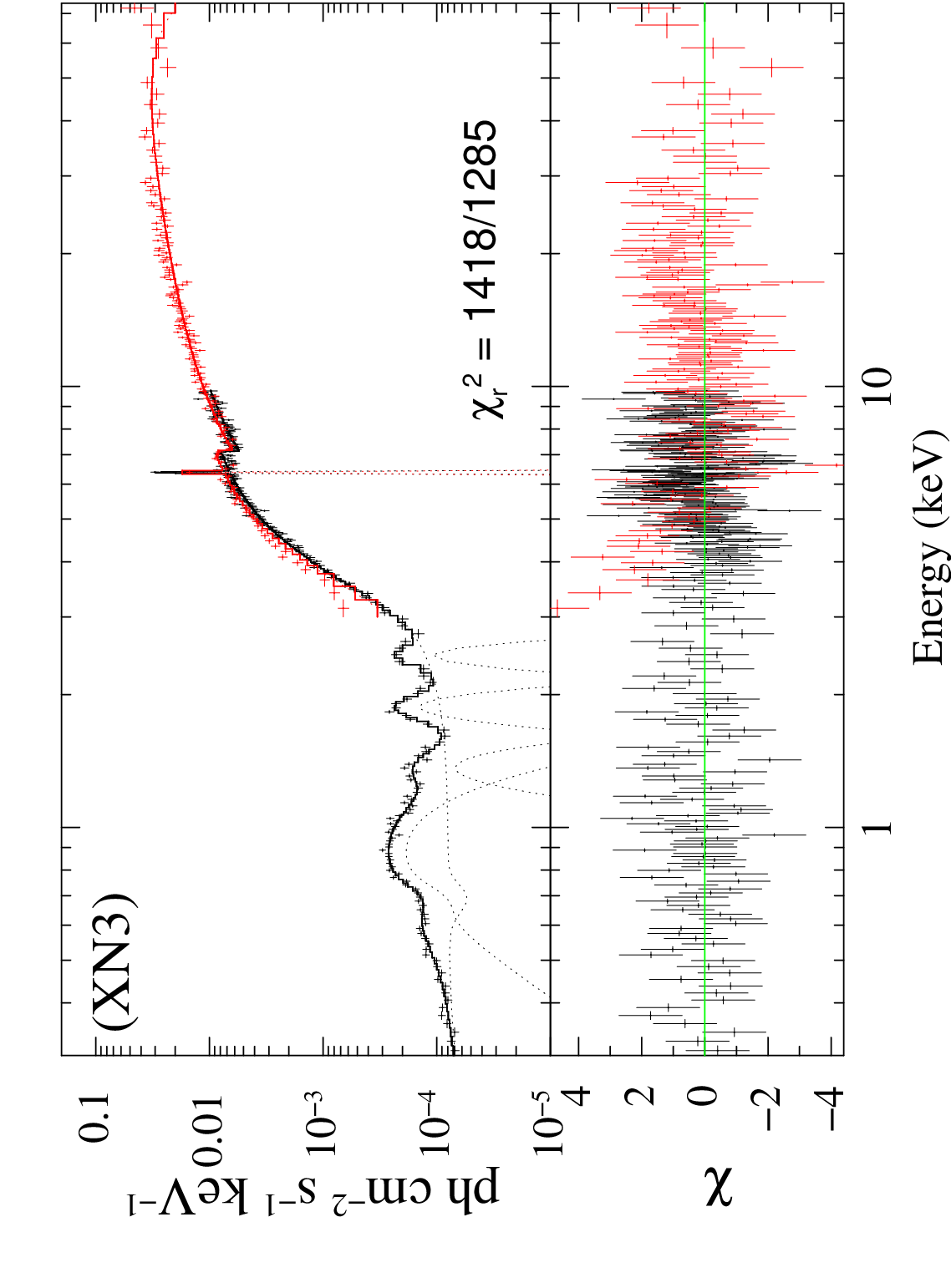}}
\caption{The \jetcaf\,model fitted spectra for all the epochs of NGC 7582. The solid and dotted lines represent the total model and the \gauss\,model components. The bottom panel shows the residuals of the fit. The fit statistics and the corresponding epochs are marked in each plot.
} 
\label{fig:specFitJoint}
\end{figure*}

The 0.3-75 keV spectra obtained in the last fifteen years are then fitted using \jetcaf\,model along with absorption components. Fig. \ref{fig:specFitJoint} shows the \jetcaf\,model fitted spectra for all the epochs of NGC 7582. We have also observed that the spectra at higher energy are not a simple power law, but rather show some excess, which has been taken care of by the bulk motion Comptonization effect in \jetcaf\, model. All model parameters are shown in Table \ref{table:JetcafResults}. Some epochs returned $\chi^2/dof\gtrsim1.2$, which can be due to the multiple peaks in the data.  \jetcaf\,is a physical model which computes continuum spectrum (no emission line) in each iteration in {\sc XSPEC} directly for a set of parameters by solving radiative transfer equations. Adding more Gaussian components may improve the fit statistics, but does not alter the \jetcaf\,model parameters. Therefore, this approach is more direct than any other phenomenological models, and the estimated model parameters are robust. The errors in \jetcaf\,parameters are estimated for a single parameter of interest, and the other independent parameters are frozen to their best fit value. We note that, since most spectra are limited up to 10 keV, \tcaf\,model equally provides good fits for them, with a marginal change in fit statistic for the {\it NuSTAR} spectra.

\begin{figure} [t!]
\hspace{-0.5cm}
\includegraphics[height=11.0truecm,width=9.0truecm,trim={0.2cm, 0.8cm, 2.0cm, 2.5cm}, clip]{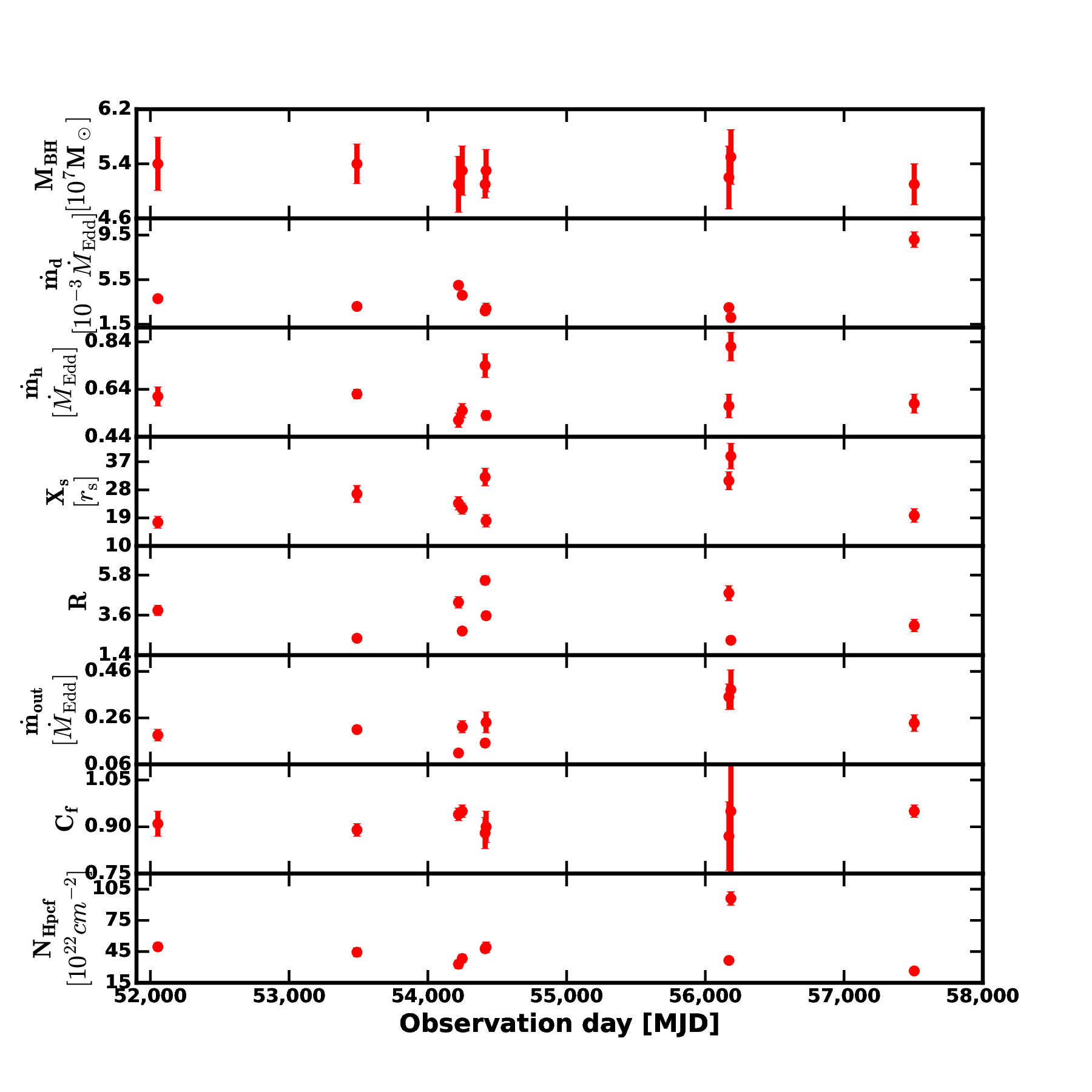}
	\caption{Variation of the fitted \jetcaf~model parameters with observational epochs. From top to bottom, panels represent variation of BH mass, disk and halo mass accretion rate, size of the corona or the location of the shock, shock compression ratio, mass outflow rate, partial covering factor and the Hydrogen column density with observation day in MJD. The bottom two rows represent the parameters of the \pcfabs\,model.}
\label{fig:FittedPars}
\end{figure}

Figure \ref{fig:FittedPars} (top-to-bottom panels) shows the temporal variation of the \jetcaf\,model parameters. The top panel shows the central AGN mass ($M_{BH}$), with an average of $5.3\times10^7$ M$_\odot$, obtained from \jetcaf\,model. The second and third panels represent the disk $(\dot m_d)$ and halo $(\dot m_h)$ mass accretion rates. The $\dot m_d$ varied significantly from $0.003$ to $0.01$ $\dot M_{Edd}$ and $\dot m_h$ from 0.47 to 0.82 $\dot M_{Edd}$. The $\dot m_h$ indicates that the hot sub-Keplerian halo component of the flow dominates over the cold Keplerian component. In case of AGN, such accretion rate behavior was often observed \citep[see][]{MandalChakrabarti2008ApJ...689L..17M,Nandietal2019,MondalEtal2022A&A...662A..77M,NandiEtal2026MNRAS.546ag072N}. That may further support that the sub-Keplerian flow is crucial in explaining black hole observations, as was originally proposed in \tcaf\,model \citep[][for a review]{ChakrabartiTitarchuk1995}. That said, the orders of magnitude higher value of $\dot m_h$ compared to $\dot m_d$ is natural, and the supply of matter is mostly from the low angular momentum winds. Such kind of accretion behavior is analogous to the hard or intermediate spectral states in low mass X-ray binary systems in \tcaf\,model scenario \citep[][and references therein]{ChakrabartiTitarchuk1995,Mondaletal2014,DebnathEtal2015MNRAS.447.1984D,ChatterjeeEtal2024ApJ...977..148C,EzeRomanus2026JHEAp..5000497E}. However, in low mass X-ray binaries the orders of magnitude difference in mass accretion rates have not been observed so far. Due to the presence of already formed Keplerian component and accretion from Roche lobe overflow, the disk mass accretion rate can be significantly high \citep{Debnathetal2014,Mondaletal2014,DebnathEtal2015MNRAS.447.1984D}.

In \tcaf\,or \jetcaf\, model scenario, the hot halo component forms the corona and therefore, the size of the corona varies with mass accretion rates. The spectrum is mostly dominated by hard photons in the flow configuration where $\dot m_h$ is higher than $\dot m_d$. However, an increase in $\dot m_d$ increases the interception of more soft photons by the hot corona, which cools it down. As a result, shock moves inward and corona shrinks \citep{MolteniEtal1996ApJ...457..805M,Mondaletal2015,ChakrabartiEtal2015MNRAS.452.3451C}. Similar behavior is also observed for NGC 7582, due to the variation in $\dot m_d$ and $\dot m_h$. The size of the corona has also changed significantly from 18 to 39 $r_S$ during the observation period (see the fourth panel of Fig. \ref{fig:FittedPars}). Here $r_S$ is the Schwarzschild radius in $2GM_{\rm BH}/c^2$, where $G$ is the universal gravitational constant, $c$ the speed of light, and $M_{\rm BH}$ is the black hole mass. The shock compression ratio parameter (R) varied in a range from 2.1 to 5.5, shown in the fifth panel of Fig. \ref{fig:FittedPars}. The estimated R values fall within the prone-to-outflow regime \citep[2-3;][]{Chakrabarti1999}, further supporting the presence of mass outflow in NGC 7582. The outflow collimation factor ($f_{\rm col}$) shows a moderately high value, indicative of isotropic outflow. Since the inflowing matter gets geometrically compressed and heated at the centrifugal barrier, some fraction of it comes out as outflows from the shock-compressed region. Therefore, the outflow rate is directly related to $R$ and $f_{\rm col}$ of the gas at the shock. This outflow remains isothermal due to radiative momentum deposition. The isothermal mass outflow rate ($\dot m_{\rm out}$) estimated \citep[using Eq. 15 in][]{Chakrabarti1999} from the combined variation of $R$ and $f_{\rm col}$ varies in a range between $0.08-0.39$ $\dot M_{\rm Edd}$, shown in sixth panel of Fig. \ref{fig:FittedPars}. This region can also be one of the ionized absorption layers associated with NGC 7582, which may influence the observed CL behavior. That said, along with the \pcfabs\, which is a weakly-ionized or neutral absorber, both can be a part of the multiphase absorber. We note that the \jetcaf\,model parameters showed a significant variability on a timescales of weeks to month. Similar timescales have been observed for other CLAGNs also; NGC 1365 showed variability on a timescale of a few hours to days \citep{Risalitietal2009}, NGC 4151 showed variability on a timescale of a few days \citep{Puccettietal2007} and NGC 1566 showed on a timescale of weeks to months \citep{XuEtal2024Univ...10...61X}. Therefore, our estimated timescales are in accord with other CLAGNs, which further connects the significant variability with the physical model parameters.

A Gaussian component is required for all epochs to fit the Fe K$\alpha$ line $\sim 6.4$ keV. This component appears from the reflection of the central continuum by the disk material \citep{Magdziarz1995,Zdziarskietal1999,YaqoobMYTORUS2012MNRAS.423.3360Y}. The Fe K$\alpha$ line is expected to be accompanied by a corresponding reflection continuum. As the continuum is reprocessed by the intervening medium, the reprocessed X-ray spectrum is characterized by significant continuum curvature peaking between $\sim$10 and 50 keV, and often a strong fluorescent Fe K emission line.

In addition, multiple Gaussian components below 2 keV are required to achieve the best fit, likely representing the so-called soft X-ray excess, a common feature in the low-energy spectra of AGN. This excess has been attributed to warm Comptonization in a warm, optically thick corona \citep{PetrucciEtal2018A&A...611A..59P}, blurred ionized disk reflection \citep{BoissayEtal2016A&A...588A..70B}, or complex partially ionized absorption imprinting atomic opacities in the soft band \citep{GierlinskiEtal2006MNRAS.371L..16G}. \cite{BianchiEtal2009ApJ...695..781B} described these peaks as emission lines from photoionized plasma in NGC 7582. We have also considered lines at 1.3 and 1.8 keV, which are identified as Mg XI and Si XIII lines by \citet{BraitoEtal2017A&A...600A.135B}, and fitted them with \gauss\, components. We note that considering the above two lines did not significantly improve $\chi^2/dof$ (<1.2) for the epochs X1, S3, and S4, which may be due to some outlier points above 2 keV. The physical origin or nature of these peaks is not yet well understood.

The epoch N2 showed another possible broad Fe K line at energy $7.78\pm0.12$ keV of width $0.36\pm0.14$ keV, with 3.2 $\sigma$ confidence estimated from F-test probability. The emission feature detected at $7.78\pm0.12$ keV is consistent with blueshifted Fe K$\beta$ emission and Ni K$\alpha$. In moderately ionized iron, the K$\beta$ transition is expected to shift toward higher energies and may blend with nearby Fe XXV satellite lines and Ni K$\alpha$, resulting in an apparent centroid around 7.7–7.8 keV \citep{MolendiEtal2003MNRAS.343L...1M,YaqoobEtal2007PASJ...59S.283Y, MondalEtal2024A&A...691A.279M}. The absence of the expected Fe K$\alpha$ line between 6.6 and 6.7 keV may suggest the intrinsic weakness of the line. Furthermore, a \gauss\, component is used for the N2 fit in the 5-6 keV energy range, which sits just below the classic Fe K$\alpha$ line at 6.4 keV. In many sources (AGN and X-ray binaries), the relativistic effects near the BH may broaden the line, which can produce a red wing extending down to $\sim 5$ keV \citep[][and references therein]{Iwasawaetal1996, ZoghbiEtal2012MNRAS.422..129Z, Milleretal2002ApJ...570L..69M, Mondaletal2016}. Such features occasionally appear as bumps rather than as sharp lines.

The absorption of the central continuum was significant along the LOS as reflected in both the \pcfabs\,model fitted parameters, $C_f$ and $N_{Hpcf}$ in panels seven and eight. The $N_{\rm Hpcf}$ value changed significantly by a factor of $\sim 4$ during the observation period, which covers the Compton thin to thick transitions as well. The $N_{\rm Hpcf}$ changed from $3.6\times10^{23}$ cm$^{-2}$ to $9.8\times10^{23}$ cm$^{-2}$ in a timescale of 15 days in 2012. This timescale is closely consistent with the optical H$\alpha$ line width changed by a factor of 4 in a timescale of 20 days during 1998 \citep{AretxagaEtal1999ApJ...519L.123A}. These timescales may help understand the short-term CS and CO behaviors if we have continuous observations.

\begin{figure}
\centering{
\hspace{-0.7cm}
\includegraphics[height=6.5truecm,width=9.0truecm,trim={1.7cm, 0.4cm, 2.0cm, 2.0cm}, clip]{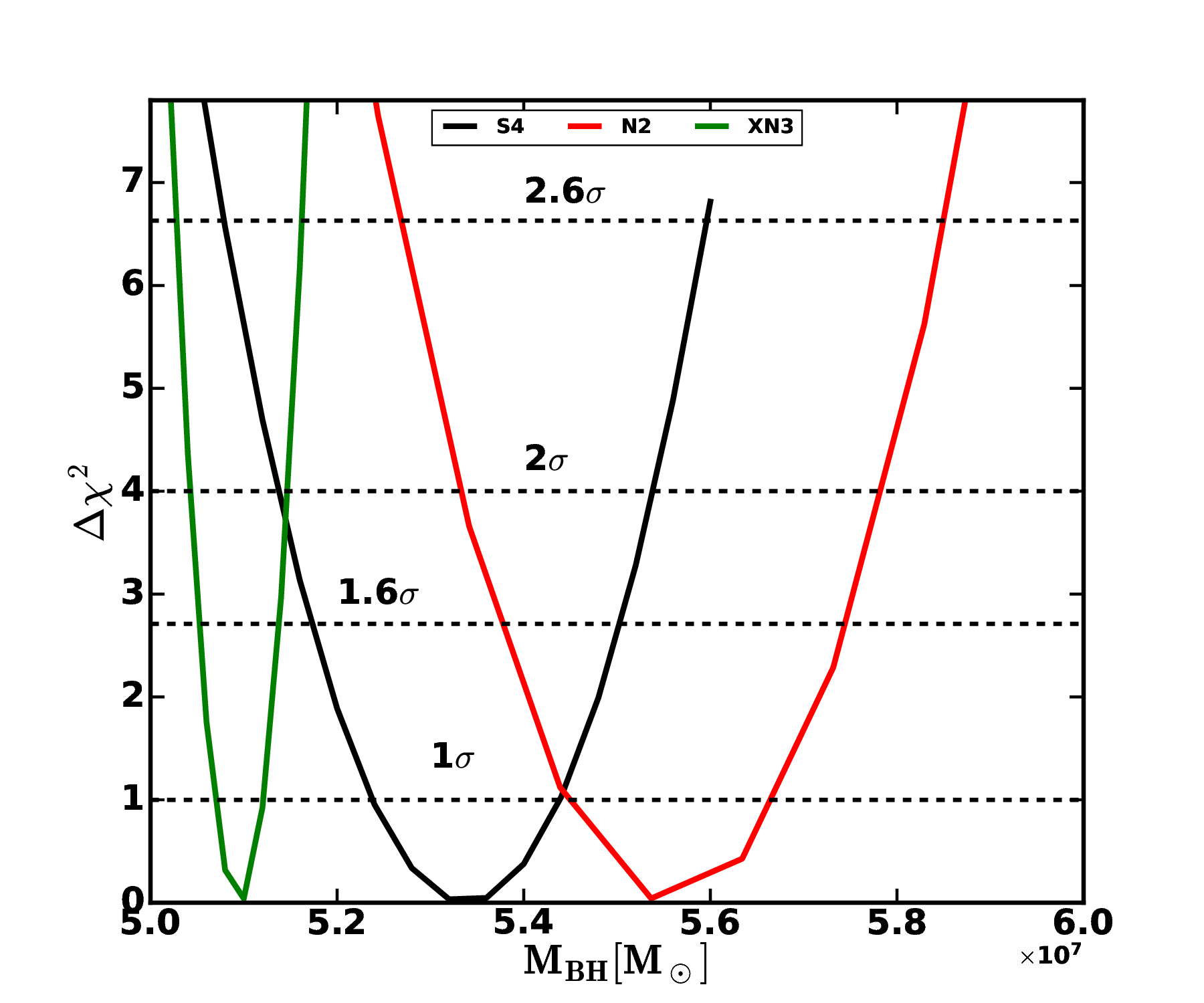}}
\caption{Different confidence levels of the estimation of SMBH mass from different epochs of data. The black, red, and green contours represent epochs S4, N2, and XN3, respectively. 
} 
\label{fig:MassConfidence}
\end{figure}

\subsection{Black hole mass}
The \jetcaf\,model has $M_{BH}$ as a user-defined parameter. Therefore, it has the capability of estimating the BH mass from the spectral modeling by keeping it free (within a range of $5\times 10^6-5\times10^9 M_\odot$) from epoch to epoch \citep[see also,][]{MollaEtal2017ApJ...834...88M,Nandietal2019,MondalEtal2022A&A...662A..77M,PandeyEtal2025A&A...695A.120P}. The $M_{BH}$ parameter varies between (5.1--5.5)$\times 10^7 M_\odot$ and the average value comes out to be $5.3 \times 10^7 M_\odot$. The estimation of $M_{BH}$ for a different significance level is also shown in Fig. \ref{fig:MassConfidence} in the $\Delta \chi^2$-$M_{BH}$ variation. The combined results from all epochs report the BH mass in NGC 7582 to be $5.3^{+0.5}_{-0.3}\times10^7 M_\odot$ in $2\sigma$ confidence level. We note that the mass estimation is done for all observations, as shown in \autoref{table:JetcafResults}. However, the confidence contours are shown for two extreme local minima of the mass estimate (green and red lines) and one from {\it Suzaku} observation (black line), which covers the robust estimates from all three satellites' data. The local minima of the mass estimates from the rest of the observations fall in between two extremes; we have not shown them in the plot. 

Different estimates of the black hole mass of NGC 7582 span a broad range of values obtained using various data sets from different instruments. Using high-resolution optical and mid-infrared observations, \citet{WoldEtal2006A&A...460..449W} estimated a BH mass in the range of $3.6-8.1 \times 10^7$ M$_\odot$ for the first time. \citet{McKernanEtal2010MNRAS.407.2399M}  calculated the BH mass $5.5\times10^{7} M_\odot$ using BASS data. \citet{RiffelEtal2009MNRAS.393..783R} estimated the BH mass $7\times10^{7} M_\odot$ using the stellar velocity dispersion. Our $M_{BH}$ agrees with the above estimates.

\begin{figure*}
\hspace{-0.6cm}
\includegraphics[height=6.0truecm,trim={6.0cm, 0.0cm, 0.0cm, 1.4cm}, clip]{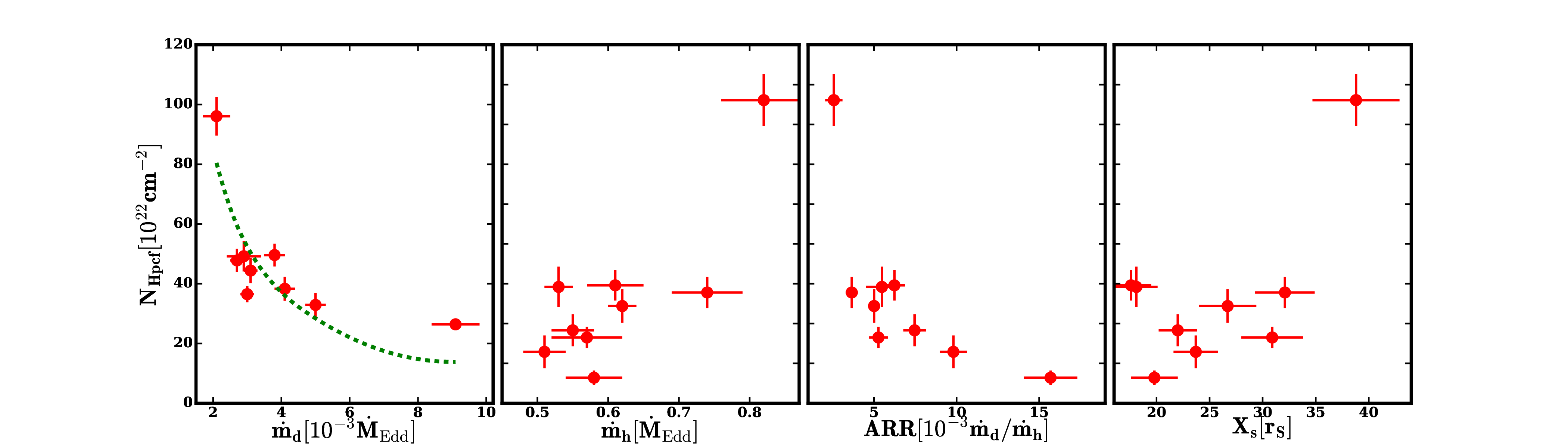}
\caption{Change in $N_{Hpcf}$ with accretion flow parameters $\dot m_d$, $\dot m_h$, $ARR$, and $X_s$ from left to right. The green dashed line shows the functional dependence of $N_H$ with $\dot m_d$.
 }
\label{fig:ScalingFittedPars}
\end{figure*}

\subsection{Origin of CL behavior: failed wind scenario}

Studies of the environment of this source reported multiple layers at different length scales with different column densities; the BLR region is one of them. The radiation coming from the central BH can help launch winds from the BLR, which can behave as an absorbing cloud along the LOS when its velocity is less than its escape velocity, i.e., the failed wind \citep{CzernyHryniewicz2011,MondalEtal2022A&A...662A..77M}. Such a failed wind physical scenario can explain the CL behavior, which is mostly governed by the underlying mass accretion. Fig. \ref{fig:ScalingFittedPars} shows the variation of the $N_{Hpcf}$ with different model parameters. The left-to-right columns show that the $N_{Hpcf}$ decreases when $\dot m_d$ increases and increases when $\dot m_h$ increases. A similar decreasing profile is observed with their ratio (which is called the accretion rate ratio or ARR). This ARR is the relative variation of the cold (Keplerian disk) and hot (sub-Keplerian hot corona) components of the accretion flow \citep{Mondaletal2014}. The $N_{Hpcf}$ shows an increasing trend with the size of the corona. Pearson correlation ranks between $N_{\rm Hpcf}$ and other accretion flow parameters in Fig. \ref{fig:ScalingFittedPars} are -0.61, 0.80, -0.91, and 0.63 for left to right panels. Similar variations of $N_{Hpcf}$ with accretion flow parameters were also observed in the case of CLAGN NGC 1365 \citep{MondalEtal2022A&A...662A..77M}. From the spectral fitting of NGC 7582, we found that $N_{\rm Hpcf}$ follows a powerlaw profile with a relation $N_{\rm Hpcf}/10^{22}$ cm$^{-2}$ = $\mathcal{N} (\dot m_d/0.001\dot m_{\rm Edd})^{-\beta}$ $\approx 0.05\pm0.02 \dot m_d^{-1.2\pm0.3}$. Here, $\mathcal{N}$ is the normalization of the fitted function. The green dashed line in the left panel of \autoref{fig:ScalingFittedPars} shows this functional dependence.

To understand the CL behavior, we have estimated different physical quantities from the spectral model fitted parameters. The radius of the BLR is estimated using the scaling relation \citep{CzernyHryniewicz2011}: $\log R_{\rm BLR}=1.538\pm0.027 + 0.5\log L_{44, 5100}$, where $R_{\rm BLR}$ is in light days and $L_{44,5100}$ is the monochromatic luminosity at 5100 \AA ~($\lambda L_{\lambda}$) measured in units of $10^{44}$ erg s$^{-1}$. The $L_{44,5100}$ can be translated to the bolometric luminosity using the relation $L_{\rm bol}=\kappa_{5100} L_{\rm 5100}$, where $L_{\rm bol}$ is calculated from $\eta \dot m_d c^2$ considering $\eta=0.1$ and $\kappa_{5100}$ is the bolometric correction factor. For local Seyfert galaxies or low luminosity AGN $\kappa_{5100} \approx 6-10$ is typically adopted \citep{KaspiEtal2000ApJ...533..631K,Netzer2019MNRAS.488.5185N}, while higher accretion-rate AGN or distant quasars may require $\kappa_{5100} \gtrsim 10$ \citep{Richardsetal2006}. For our present analysis, we have adopted $\kappa_{5100}=10$. However, considering the lower value of 6 only marginally changes the results.
Given the spectral model fitted $\dot m_d$ from Table \ref{table:JetcafResults}, the $R_{\rm BLR}$ ranges between $3.4-6.8\times10^{15}$ cm $\sim 210-420~r_S$.
In the same line, estimating the BLR radius using different scaling relations \citep{KaspiEtal2005ApJ...629...61K,BentzEtal2013ApJ...767..149B} may shift the values by a factor of 2-3, but that does not alter the main interpretation of the work.  

\begin{figure}
\hspace{-0.5cm}
\includegraphics[height=5.6truecm,trim={0.0cm, 0.0cm, 0.0cm, 0.0cm}, clip]{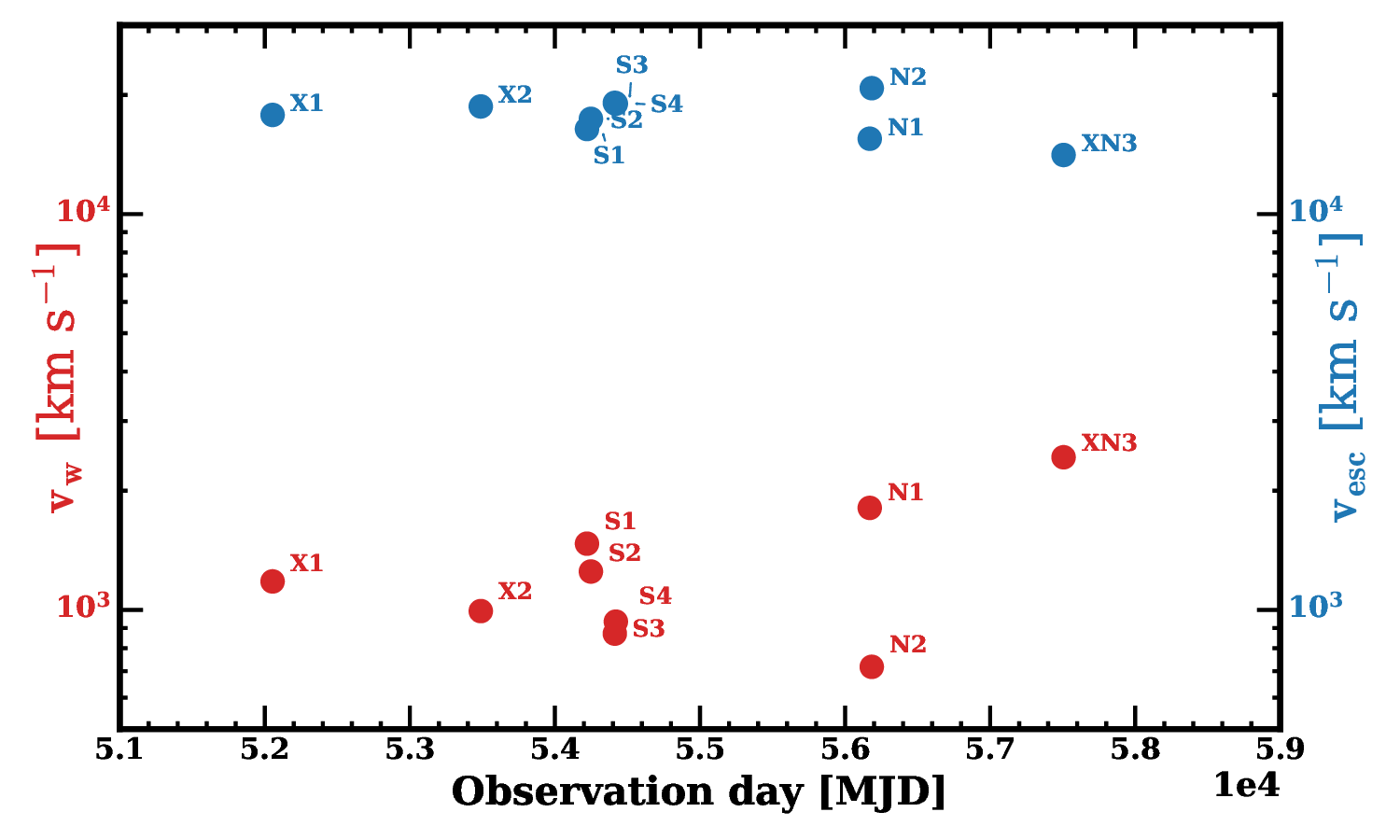}
\caption{Velocity profiles of the wind launched from the disk with observation time. The red (left-axis) and blue (right-axis) circles denote the wind and escape velocities. The observation epochs are marked with each data point. Both velocities are plotted on a logarithmic scale.
 }
\label{fig:windVelocity}
\end{figure}

Next, we model the wind as a partial, thin spherical shell at a distance $R_{\rm BLR}$ from the central source, moving outward from the accretion disk with a constant velocity $v_{\rm w}$. The resulting mass outflow rate is given by
$\dot M_{\rm w}=\mu m_p N_{\rm H} v_{\rm w} R_{\rm BLR} \Omega,$ 
where $\mu=1.4$, $m_p=1.67\times 10^{-24}$~gm, and $\Omega=\pi/2$ \citep{Maiolino1995} are the mean atomic mass per proton, proton mass, and the typical value of the solid angle subtended by the outflow, respectively. Assuming that the radiation from the central source is incident onto the wind, and each emitted photon scatters approximately once before escaping to infinity, momentum conservation implies that the total wind momentum rate is of the order of the photon momentum flux, $\dot p_w=\dot M_w v_w \sim L_{\rm bol}/c$ \citep{King2010} and the wind acceleration is due to electron scattering. Using the above relation, we derive an estimated wind velocity  \citep{MondalEtal2022A&A...662A..77M}, $v_{\rm w}=30 \left(L_{\rm bol}\dot m_{\rm d}^\beta/\mathcal{N} R_{\rm BLR}\right)^{1/2}$ km s$^{-1}$.  Given the estimated $L_{\rm bol}$, $\mathcal{N}$, $\dot m_d$, and $R_{\rm BLR}$, the wind velocities for the epochs X1, X2, S1-S4, N1, N3, and XN3 come out to be 1180, 993, 1470, 1250, 871, 934, 1810, 718, and 2430 km s$^{-1}$, respectively. This estimate is consistent with the lower limit of the velocity (which is 637 km s$^{-1}$) of the obscuring cloud reported by \citet{LefkirEtal2023MNRAS.522.1169L}. Furthermore, we have also estimated the escape velocity ($v_{esc}$) of the wind from the BLR region using $(2 G M_{BH}/R_{BLR})^{1/2}$ km s$^{-1}$. Both velocities are compared in Fig. \ref{fig:windVelocity}. The red and blue points with observation epochs labels denote the $v_w$ and $v_{esc}$ respectively. It is evident from the plot that $v_w$ is much less compared to $v_{esc}$, therefore, wind launched from the disk could not escape the gravitational potential of the SMBH. This failed wind might be one of the multiple absorbers that obscured the central radiation and originated the CL behavior in NGC 7582. {\color{black}In this analysis, the photon momentum is assumed to be uniquely transferred to drive wind with no other means such as thermal pressure and magnetic force, which could result in an uncertainty in the current estimate on the wind velocity, even if indeed launched.}

The above failed wind {\color{black}or obscurer} scenario is formulated using the \pcfabs\,model parameter $N_{\rm Hpcf}$. Though the \pcfabs\,model is a phenomenological continuum absorption model rather than an ionic line-resolved photoionization model (such as \textsc{cloudy} or \textsc{xstar}) and the $N_{\text{Hpcf}}$ is constrained via photoelectric continuum absorption rather than individual ionic absorption lines (e.g., Ne X, Si XIV, Fe XXVI), it directly probes the cool, high-density core phase of the outflowing obscurer. The photoelectric opacity of this component is determined by the combined cross-sections of metals such as O, Ne, Mg, Si, S, and Fe, assuming standard abundances \citep[e.g.,][]{Wilmsetal2000}. Thereby, the continuum curvature and edge depths trace the total integrated equivalent $N_H$. The physical association of this neutral/partially ionized gas with an outflowing disk wind {\color{black}or obscurer} of characteristic velocity $v_{\rm w}$ {\color{black}can be} justified through the rapid time-variability in $N_{\text{Hpcf}}$. {\color{black}We note that, the measurement of obscuring column in conjunction with its partial covering variability is used as a proxy to phenomenologically imply partial property of winds/outflows without detection of absorption lines in the spectrum.}

Our study shows that the failed wind scenario is associated with the low disk accretion rate, where the outflowing wind can block the central radiation potentially. Therefore, any small fluctuation in accretion rate can appear as a significant variation in the observed spectra. Additionally, instabilities may grow in the disk as the wind is launched and reined back. As a result, the BLR can form or disappear on short timescales, leading to dramatic changes in optical spectral type. Together, these effects provide a natural explanation for why CL behavior is more likely to occur at a low disk accretion rate. This further refers to our earlier discussion that the corona is dominating the radiation ($\dot m_h> \dot m_d$) and playing a crucial role in spectral variability.

Since other mechanisms can also launch winds from the accretion disk, we are considering the current assumption as a possibility and therefore the origin of the failed winds. While line-driven radiative acceleration \citep[e.g.,][]{ProgaEtal2000ApJ...543..686P} is commonly used for AGN disk winds, the magnetic driving via magnetocentrifugal forces \citep[see,][]{FukumuraETal2010ApJ...715..636F} offers a highly efficient alternative or complementary acceleration mechanism capable of launching dense, low-ionization gas beyond the local escape velocity ($v_{\text{esc}}$). Although our continuum absorption analysis using \pcfabs\,model does not provide explicit line blueshift kinematics, the derived $N_{\text{Hpcf}}$ and $C_f$ decide the properties of the obscuring material quantitatively. These physical boundary conditions are essential for testing both magnetohydrodynamic and radiatively driven multi-phase wind models in sub-Eddington AGNs. This remains an open question and warrants further investigation in future studies.

\subsection{Timescales of CL transitions}
NGC 7582 has undergone several state transitions on different timescales: X-ray CO transition in over $\sim$15 days in 2012, optical Seyfert 2$\rightarrow$Seyfert 1 over $\sim$20 days in 1998, and X-ray CO transition in over $\sim$6 months in 2007, along with possibly three Compton-thick/Compton-thin transitions between 2005 and 2014 \citep[see Sec 1 and ][]{LefkirEtal2023MNRAS.522.1169L}. Therefore, it is clear that no single mechanism could produce such diverse timescales; rather, different processes are associated with different timescales. 

Since the change in viscosity can change the mass accretion behavior, therefore, the role of viscosity is one of the plausible and promising mechanisms \citep{Shakura1973,ChakrabartiBook1990ttaf.book.....C,Chakrabarti1996ApJ...464..664C,Mondaletal2017}. The viscosity timescale ($t_\nu$) for AGN is much longer, $\sim$ a few hundred years to a few Myr when the mass is accreted from the outer disk \citep[][and references therein]{NodaDone2018,MondalEtal2022AA...663A.178M,Ricci2023NatAs...7.1282R,HagenEtal2026A&A...708A...5H} and scales with the central BH mass. However, growing instabilities due to hydrostatic or magnetic effects in an accretion disk may considerably change $t_\nu$ \citep{JaniulEtal2011MNRAS.414.2186J,DexterBegelman2019MNRAS.483L..17D}. In addition, several physical mechanisms may contribute to the growth of such instabilities, including thermal instability in an accretion disk \citep{LinShields1986}, propagation of cooling front outward through the accretion disk and driven by changes in magnetic torque at the inner disk \citep{SternEtal2018CLQ}, interaction of an accretion disk tidally with binary star \citep{Riccietal2020}. The characteristic timescales of these processes are broadly consistent with the observed variability timescales of CLAGN, particularly during state transitions \citep[][for a review]{Ricci2023NatAs...7.1282R}. Furthermore, the wind launched from the disk may carry both the energy and angular momentum of the flow. It can also enhance instabilities in the disk and substantially reduce the viscous timescale \citep[][and references therein]{SniegowskaEtal2020A&A...641A.167S,PanEtal2021ApJ...910...97P}. Hence, $t_\nu$ can vary in a broad range depending on the region of instability, the amount of mass accreted, and the viscosity present in the disk. This underscores the need for estimating different timescales in the accretion disk of NGC 7582.

Given the best-fit spectral model parameters, we have estimated $t_\nu$ for NGC 7582. Since the wind can be launched from anywhere on the disk, e.g., either from the inner corona or from the outer BLR region, we have estimated $t_\nu$ for both regions of the disk.  We use the relation $t_\nu=t_{\rm dyn} (H/r)^{-2} \alpha$, where $t_{\rm dyn} = (G M_{BH}/r^3)^{-1/2}$ is the dynamical timescale of the infalling matter and $\alpha$ is the viscosity parameter. The scale height of the disk $(H)$ in the BLR region is given by $H_{BLR} = 3 \dot m_d \kappa/8\pi c$, assuming that $\dot m_d$ is constant throughout the disk, where $c$ is the speed of light. This expression is similar to \citet{Shakura1973} solution with opacity $\kappa = 50$ cm$^2$ gm$^{-1}$ for the BLR region dominated by the dust \citep[see][]{BaskinLaor2018MNRAS.474.1970B}. We note that the above relation can also be scaled with the bolometric luminosity, which varies significantly from one CLAGN to the other. Thus, the disk aspect ratio $H/R$ also varies in a broad range, in contrast of being a constant for CLAGN of mass ranges from $10^6-10^{8.5} M_\odot$. Therefore, $t_\nu$ also varies significantly by some orders of magnitude, which prompted us to calculate $H$ at different regions of the disk at different epochs. As the disk puffs up in the inner hot corona region, the scale height is estimated using the relation $H_{shk}=[\gamma (R-1)X_s^2/R^2]^{1/2}$ \citep[Eq. 1 in][]{Debnathetal2014}. For the estimated $R_{BLR}$ and spectral model fitted $X_s$ above, the $t_{dyn}$ yields in a range of $\sim 26-79$ days and $\sim 0.7-2.2$ days, respectively. The estimated $t_\nu$ is $>10^5$ years for the BLR region, which is significantly longer compared to the observed state transition timescale and $\sim 20-60$ days for the corona region. Furthermore, we have estimated the thermal timescale ($t_{th}=t_{\rm dyn}/\alpha$), which is $\sim 8$ months to $\sim 1.7$ years for the BLR region, while it is $\sim 7-22$ days for the corona region. Throughout the analysis, we used $\alpha=0.1$, which may not be the case for the disk in all epochs; rather, it should change with time. Since we do not have its estimation beforehand, we considered a standard value. However, a choice of lower $\alpha$ will return higher timescales and vice versa. In addition, considering the fluctuation of a tiny region of the BLR may reduce $t_\nu$ sufficiently as formulated in \citet{SniegowskaEtal2020A&A...641A.167S}, but not to the extent that it satisfies the CL timescale observed to date.

\begin{figure}
\hspace{-0.4cm}
\includegraphics[height=5.5truecm,trim={0.0cm, 0.0cm, 0.0cm, 0.0cm}, clip]{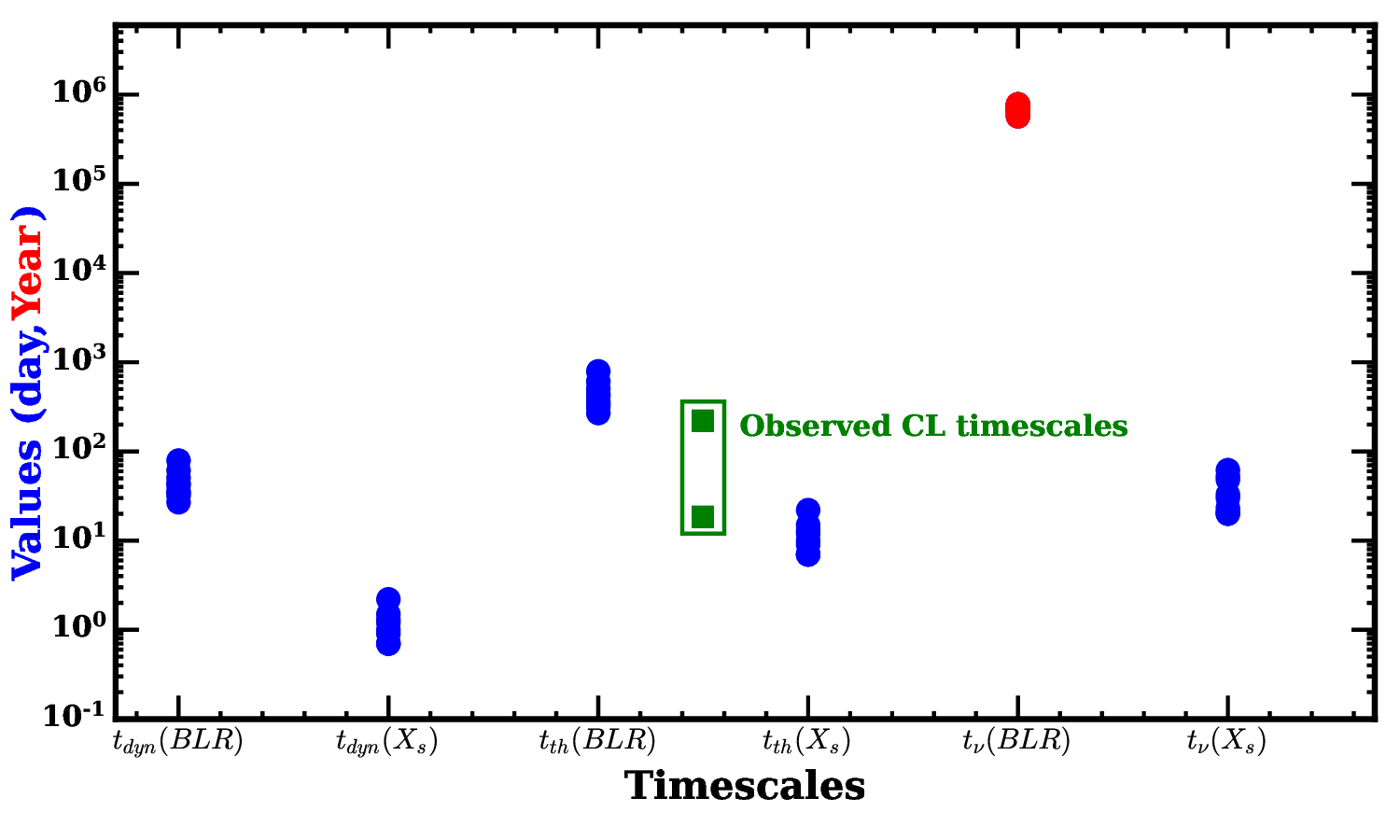}
\caption{Estimated timescales of spectral state transition in NGC 7582 from different disk regions. Blue and red circles represent timescales in days and years. The green rectangle region denotes the observed timescales of state transition as mentioned in the text. The values of timescales are plotted on a logarithmic scale.
 }
\label{fig:timescales7582}
\end{figure}

In Fig. \ref{fig:timescales7582}, we show three different timescales (see Table \ref{table:Timescale} as well) estimated for both the BLR and corona regions. All blue circles denote timescale values in day units, whereas the red circles denote the timescale in year units. The green points represent the observed timescales of the spectral state transition in NGC 7582. Our combined estimations of different timescales show that the observed CL transitions for this source can be explained by the variability in both corona and BLR regions. The dynamical timescale at the BLR region is short, close to the shortest observed CL transition in X-ray in 2012 and to some extent in optical in 1998. This associates the role of BLR region and the possibility of the failed-wind scenario in dynamical timescales. On the other hand, the viscous and thermal timescales near the corona region are also consistent with the observed CL transitions. Therefore, the observed relationship between $\dot{m}_d$ and $N_H$ suggests that the changes in the accretion flow regulate the column density via variability in the corona. If it occurs on timescales comparable to, or shorter than, those associated with the CL behavior (e.g., variations in optical line widths and/or $N_H$), the BLR then responds on its dynamical timescale and thereby contributes to the observed CL behavior. This study thus highlights a direct physical link between the corona and the BLR, which may be potential drivers of the CL behavior.

\begin{table*}
\scriptsize
\centering
	\caption{\label{table:Timescale} Calculated dynamical ($t_{\rm dyn}$), thermal ($t_{\rm th}$), and viscous ($t_\nu$) timescales for the BLR and sub-Keplerian shock region ($X_s$) across observation epochs.} 
\begin{tabular}{ccccccc}
\hline
Epoch & $t_{\rm dyn} (\text{BLR})$ & $t_{\rm dyn} (X_s)$ & $t_{\rm th} (\text{BLR})$ & $t_{\rm th} (X_s)$ & $t_\nu (\text{BLR})$ & $t_\nu (X_s)$ \\
      &(days)& (days)&(days) &(days)& (year)&(days)\\
\hline
X1  & 41.7 & 0.7 & 417 & 7.0 & $6.8 \times 10^5$ & 21.0 \\
X2  & 35.8 & 1.2 & 358 & 12.0 & $7.2 \times 10^5$ & 30.0 \\
\hline
S1  & 50.5 & 1.0 & 505 & 10.0 & $6.6 \times 10^5$ & 33.0 \\
S2  & 43.9 & 0.9 & 439 & 9.0  & $6.8 \times 10^5$ & 23.3 \\
S3  & 31.8 & 1.5 & 318 & 15.0 & $7.7 \times 10^5$ & 61.7 \\
S4  & 33.9 & 0.7 & 339 & 7.0  & $7.4 \times 10^5$ & 20.0 \\
\hline
N1  & 61.1 & 1.3 & 611 & 13.0 & $6.2 \times 10^5$ & 48.1 \\
N2  & 26.9 & 2.2 & 269 & 22.0 & $7.8 \times 10^5$ & 53.1 \\
XN3 & 79.1 & 0.7 & 791 & 7.0  & $5.7 \times 10^5$ & 20.1 \\
\hline
\end{tabular}
\end{table*}

\section{Conclusions} \label{sec:Conclusion}

In this work, we studied the changing look behavior in  NGC\,7582 using fifteen years of X-ray data. We performed spectral modeling using an accretion-ejection based \jetcaf\,model along with the partial absorption, \pcfabs\,models to understand the origin of CL behavior in this system. We conclude the following based on our analyses and estimations of physical quantities of accretion.

\begin{itemize}
\item The spectral modeling of long term X-ray data reveals that the disk mass accretion rate (Keplerian) is $\lesssim 0.01 \dot M_{\rm Edd}$ and the halo  mass accretion rate is $<0.82 \dot M_{\rm Edd}$. The size of the corona varied in a range between $\sim 18-39 r_S$.

\item The central SMBH mass estimated directly from spectral modeling and using the $\chi^2$ minimization method comes out to be $5.3^{+0.5}_{-0.3}\times10^7 M_\odot$.

\item An Iron line at $\sim 6.4$ keV was present in all spectra, while another Fe K line at $\sim 7.8$ keV was present on epoch N2.

\item The neutral hydrogen column density $N_{Hpcf}$ varied significantly from $26\times10^{22}$ cm$^{-2}$ to $96\times10^{22}$ cm$^{-2}$ during the entire observation period, and it decreased with disk mass accretion rate.

\item The central continuum was significantly absorbed by the gas cloud along the LOS with a partial covering factor $>0.87$.

\item While photoelectric continuum fitting ($N_{\text{Hpcf}}$) lacks line blueshifts to directly measure wind velocity ($v_w$), short-term $N_{\text{Hpcf}}$ and $C_f$ variability confirms the absorbing gas is dynamically active across the LOS. If the launched wind has velocity below the escape velocity ($v_w < v_{\text{esc}}$), this may support a ``failed wind {\color{black}or obscurer}'' scenario as the origin for the CO behavior.

\item NGC 7582 showed the CO transition in X-rays in a much shorter timescale, which can be explained either by the dynamical timescale at the BLR or by the viscous and thermal timescales at the boundary layer of the corona. Our study thus highlights a direct physical link between the corona and the BLR, which may be the potential drivers of the CO behavior, which is a sub-class of the CL behavior.

\end{itemize} 

Given the detailed spectral analysis of the long-term X-ray data and the estimation of different physical quantities, we report that the CL behavior in NGC 7582 can be explained by the failed wind scenario. Furthermore, the variation in mass accretion is the key that can solely explain the origin of CL behavior. These results are in accord with our previous study of CLAGN NGC 1365 \citep{MondalEtal2022A&A...662A..77M}, and we anticipate that other potential CLAGNs may satisfy the same physical scenario. As a further comparison with the recently observed (using {\it XRISM}) {\color{black}moderately} obscured Seyfert 1 galaxies NGC 3783 {\color{black}\citep[see][for further obscuration behavior]{MehdipourNGC37832025A&A...699A.228M}} and NGC 5548 \citep{MehdipourNGC55482026ApJ..1008...56M}, NGC 7582 showed potential similarities and differences. The obscuring gas in NGC 3783 and NGC 5548 exists as a multi-phase stream (as probed via {\color{black}photoionization calculations}). NGC 7582 shows a similar multi-phase nature, where the \pcfabs\,component traces the dense and clumped cores of the outflow. Furthermore, both NGC 3783 and NGC 5548 underwent multi-year state transformations driven by passing obscuring streams that block ionizing EUV/soft X-ray radiation. This behavior is consistent with the long-term X-ray variability and CL behaviors observed in NGC 7582. However, there are some differences which have been observed in XRISM observation, e.g., direct blueshifted absorption lines are detected for NGC 3783 and NGC 5548, providing precise LOS velocity constraints and more precise measurement of the escaping wind velocity, which is not observed in the present data sets of NGC 7582. {\color{black}In these studies \citep{MehdipourNGC37832025A&A...699A.228M,MehdipourNGC55482026ApJ..1008...56M}, the characterization of the obscurers is done by photoionization modeling to securely constrain outflow properties all together, leaving little doubt about their outflow kinematic condition, with which the authors are able to estimate proper outflow kinetic power and momentum transfer to the host galaxy.}

\authorcontributions{SM conceptualized the paper, analyzed and modeled the data, and written the paper. VJ and NK processed some of the instruments data used here. TPA went through the paper and made scientific comments. All authors contributed in writing and presenting the results.}

\funding{VJ acknowledges the support provided by the Department of Science and Technology (DST) under the ‘Fund for Improvement of S \& T Infrastructure (FIST)’ program (SR/FST/PS-I/2022/208). V.J. also thanks IUCAA, Pune, India, for the Visiting Associateship. NK acknowledges funds from European Union - Next Generation EU, Mission 4 Component 1 CUP C53D23001330006 and INAF Large Grant 2023 BLOSSOM F.O. 1.05.23.01.13. TPA acknowledges the support of the National Natural Science Foundation of China (grant nos. 12222304, 12192220, and 12192221).}

\institutionalreview{Not applicable}

\informedconsent{Not applicable}

\dataavailability{All data are publicly available in NASA's HEASARC archive.}

\acknowledgments{We thank both the Referees for their constructive comments and suggestions. This research has made use of the {\it NuSTAR} Data Analysis Software ({\sc nustardas}) jointly developed by the ASI Science Data Center (ASDC), Italy, and the California Institute of Technology (Caltech), USA. This research has also made use of data obtained through the High Energy Astrophysics Science Archive Research Center Online Service, provided by NASA/Goddard Space Flight Center. This research has made use of XMM-Newton/EPIC-pn observations, an ESA science mission with instruments and contributions directly funded by ESA Member States and NASA. This research has made use of data obtained from the Suzaku satellite, a collaborative mission between the space agencies of Japan (JAXA) and the USA (NASA).}

\conflictsofinterest{The authors declare no conflicts of interest.}


\appendixtitles{yes}
\appendixstart
\appendix

\section{Illustration of the \jetcaf\,model}\label{sec:apenA}
\begin{figure}
\includegraphics[height=7.0truecm,trim={0.0cm, 0.0cm, 0.0cm, 0.0cm}, clip]{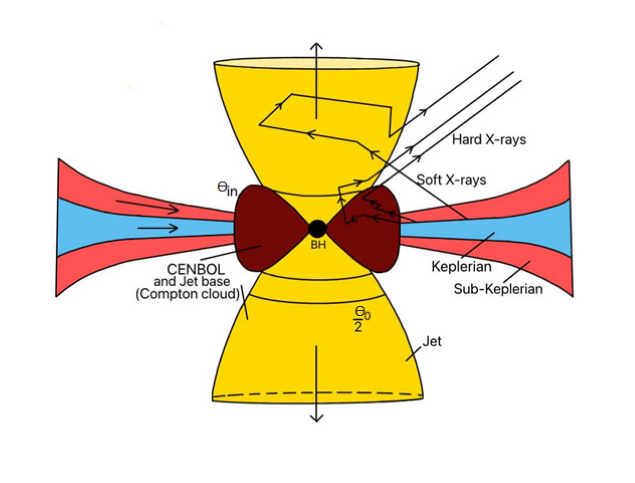}
\caption{Illustration of the \jetcaf\,model. Different components of the flow: hot (sub-Keplerian), cold (Keplerian) components, the Compton cloud (corona), and the outflowing jets are shown in different colors. The zig-zag paths show the scattering of photons by different media of the inflow and the outflow. The figure is adopted from \citep{MondalChakrabarti2021}.
 }
\label{fig:JeTCAFIllus}
\end{figure}

The \jetcaf\,model has six parameters: (i) black hole mass ($M_{\rm BH}$), (ii) disk mass accretion rate or cold Keplerian flow ($\dot m_d$), (iii) halo mass accretion rate or hot sub-Keplerian flow ($\dot m_h$), (iv) size of the corona or the location of the shock ($X_s$), (v) shock compression ratio or the density jump factor ($R$), and jet or outflow collimation factor ($f_{\rm col}$). Here, we briefly discuss how the spectral shape changes with model parameters. \autoref{fig:JeTCAFIllus} shows the illustration of the model, where the cold (blue) flow is sandwiched between the hot (red) flow. Two funnel-shaped regions (in yellow) above and below the corona (magenta) represent the jets. These corona and jet regions inverse Comptonize the soft photons from the cold disk and make them harder. The zig-zag paths show the scattering of the photon by hot electrons at different regions of the accretion-ejection flows. 

As the $\dot m_d$ increases, the number of soft photons increases, thereby both jet and corona spectra soften due to the dominance of the soft photons. An increase in $\dot m_h$, the high-energy photons contribute more, and the spectra become harder. Increase in $X_s$ and $R$ softens the jet spectrum as the temperature and mass outflow rate in the jet decrease \citep[see,][]{ChakrabartiTitarchuk1995}, while the spectrum hardens with the increase in $f_{\rm col}$. Overall, the presence of the jet hardens the spectral component at the shoulder of the blackbody component, which comes from the scattering of the soft photons by the jet medium. As the jet is moving with a velocity, the effect of jet bulk motion can also produce a hump-like feature above 10 keV. Therefore, the spectral components are: (i) power-law from the Compton cloud or corona due to inverse Comptonization of the soft photons, (ii) disk blackbody from the cold disk, (iii) another inverse-Comptonized component at the shoulder of the blackbody originates from the jet, (iv) the bulk motion Comptonized component when hard photons from the corona pass through the moving jet, and (v) reprocessing of the hard photons by the cold disk iteratively taken into account. The variation in the spectral shape with model parameters is described in detail in \citet{MondalChakrabarti2021}. Our explanation of spectral properties is also consistent with the detailed spectral studies presented in \citet{Chakrabarti1997} for accretion around black holes.

 As a note, the current \jetcaf\,model calculations do not include dissipation of energy at the shock. There can be changes in the accretion flow parameters if the shock dissipates energy \citep[see][]{DasEtal2010MNRAS.401.2053D,MondalEtal2014ApSS} such as: 1. A much higher compression ratio implies a denser, more compact corona. Consequently, the optical depth ($\tau$) of the corona would increase substantially for the identical mass-accretion rates, modifying the Comptonization spectra and reducing the required mass-inflow variations to trigger spectral state transitions, 2. The inclusion of rapid energy dissipation would likely shorten the timescale of state transitions, 3. A cooler corona produces softer spectra. This would shift the model-fitted accretion rate parameters ($\dot{m}_d, \dot{m}_h$ and shock locations ($X_s$) needed to fit specific X-ray states, likely requiring a higher mass-inflow rate, and 4. If energy is dissipated significantly, the thermal pressure driving the outflow decreases, and the shock moves inward. While centrifugal forces still operate, the mass outflow-to-inflow ratio ($R_{\dot{m}}$) and the initial jet velocity would be modulated. However, in the present analysis model parameters are robust as the spectra are fitted directly using the model, which is not a phenomenological but a physical model solving multiple fundamental equations. A further comparative study with and without dissipation may help understand such effects.

\end{paracol}
\externalbibliography{yes}
\bibliography{clagns} 

\end{document}